\documentclass[fleqn,usenatbib]{mnras}

\usepackage{newtxtext,newtxmath}
\usepackage{subcaption}
\usepackage{verbatim}

\usepackage[T1]{fontenc}

\DeclareRobustCommand{\VAN}[3]{#2}
\let\VANthebibliography\thebibliography
\def\thebibliography{\DeclareRobustCommand{\VAN}[3]{##3}\VANthebibliography}

\usepackage{graphicx}	
\usepackage{amsmath}	

\title[H{\sc i}-bearing UDGs in WALLABY]{WALLABY: an untargeted search for H\,{\sc i}-bearing ultra-diffuse galaxies uncovers the first known ultra-diffuse galaxy pair}

\author[T. O'Beirne et al.]{
T. O'Beirne,$^{1,2,3}$\thanks{E-mail: tamsynobeirne.astro@gmail.com}
V. A. Kilborn,$^{1}$
M. E. Cluver,$^{1}$
O. I. Wong,$^{2,4}$
N. Deg,$^{5}$
K. Spekkens,$^{5}$
N. Arora,$^{6,5}$
\newauthor
R. Dudley,$^{7,5}$
B. Catinella,$^{4}$
H. Dénes,$^{8,9}$
K. Lee-Waddell,$^{10,2,11}$
P. E. Mancera Piña,$^{12}$
C. Murugeshan,$^{13}$
\newauthor
J. Rhee,$^{2,4}$
L. Staveley-Smith,$^{4}$
A. X. Shen,$^{2}$
and T. Westmeier$^{4}$
\\
$^{1}$Centre for Astrophysics and Supercomputing, Swinburne University of Technology, Hawthorn, Victoria 3122, Australia\\
$^{2}$CSIRO Space \& Astronomy, PO Box 1130, Bentley WA 6102, Australia\\
$^{3}$European Southern Observatory, Karl-Schwarzschildstrasse 2, D-85748 Garching bei München, Germany\\
$^{4}$International Centre for Radio Astronomy Research (ICRAR), The University of Western Australia, 35 Stirling Highway, Crawley, WA 6009, Australia\\
$^{5}$Department of Physics, Engineering Physics, and Astronomy, Queen’s University, Kingston ON K7L 3N6, Canada\\
$^{6}$Arthur B. McDonald Canadian Astroparticle Research Institute\\
$^{7}$David A. Dunlap Department of Astronomy, University of Toronto, Toronto ON M5S 3H4, Canada\\
$^{8}$School of Physical Sciences and Nanotechnology, Yachay Tech University, Hacienda San José S/N, 100119, Urcuquí, Ecuador\\
$^{9}$Universidad San Francisco de Quito, Quito, Ecuador\\
$^{10}$Australian SKA Regional Centre (AusSRC) - The University of Western Australia, 35 Stirling Highway, Crawley WA 6009, Australia\\
$^{11}$ICRAR - Curtin University, Bentley, WA 6102, Australia\\
$^{12}$Leiden Observatory, Leiden University, P.O. Box 9513, 2300 RA, Leiden, The Netherlands\\
$^{13}$Australia Telescope National Facility, CSIRO, Space and Astronomy, P.O. Box 76, Epping, NSW 1710, Australia
}

\date{Accepted XXX. Received YYY; in original form ZZZ}

\pubyear{\the\year{}}

\begin{document}
\label{firstpage}
\pagerange{\pageref{firstpage}--\pageref{lastpage}}
\maketitle

\begin{abstract}
Using the Widefield ASKAP L-band Legacy All-sky Blind surveY (WALLABY) we performed an untargeted search for H\,{\sc i}-bearing ultra-diffuse galaxies (UDGs).  We identified a core sample of 10 UDGs defined by $\mu_{g,0}\ge24$ mag~arcsec$^{-2}$ and $R_{e}\ge1.5$~kpc, and a broader sample including 12 additional faint diffuse galaxies ($\mu_{g,0}\ge23.7$~mag~arcsec$^{-2}$ and $R_{e}\ge1.3$~kpc). Within the core sample, we highlight the first discovery of a UDG pair. Their projected separation is just 75 arcsec (22~kpc at 61.9~Mpc), with a central H\,{\sc i} velocity difference of 34 km~s$^{-1}$.  The North-Western UDG (WALLABY J104513-262755-UDG-1) has a larger H\,{\sc i} reservoir, $\log_{10}(M_{HI}/\rm M_{\odot}) = 8.95\pm0.03$, compared to the South-Eastern UDG (WALLABY J104513-262755-UDG-2), $\log_{10}(M_{HI}/\rm M_{\odot}) = 8.60\pm0.04$. UDG-1's stellar mass and star formation rate are also approximately an order of magnitude larger at $\log_{10}(M_*/\rm M_{\odot}) = 8.07\pm0.12$ and $\log_{10}(SFR/\rm M_{\odot}~yr^{-1}) = -1.26\pm0.12$ respectively. The pair has an isolated local environment, with no other galaxies or H\,{\sc i} sources within 30 arcmin (525 kpc) and $\pm1000$ km~s$^{-1}$. { However, in the context of the larger-scale structure, the pair is located outside the virial radius of the Hydra cluster, with its position on the phase-space diagram indicating that it is infalling into the cluster.}  The identification of this H\,{\sc i}-bearing UDG pair raises important questions around the formation of such a unique system and the evolution of UDGs in a transitional phase before ram pressure stripping and cluster infall.


\end{abstract}




\section{Introduction}
\label{sec:intro}


Optically low surface brightness galaxies (LSBGs) have been predicted to represent a significant galaxy population. They may make up $85\%$ of the number density for galaxies with stellar masses $>10^7$ M$_{\odot}$ \citep[][]{Martin:19}. While the study of LSBGs has a well established history \citep[e.g.][]{Bothun:87,Impey:88}, up until the recent development of sensitive optical and H\,{\sc i} surveys, such as the Dark Energy Spectroscopic Instrument (DESI) Legacy Imaging Surveys \citep[hereafter refered to as the Legacy Survey;][]{Dey:19} and the Arecibo Legacy Fast ALFA \citep[ALFALFA;][]{Giovanelli:05}, they have been incredibly difficult to detect in large numbers. Yet our understanding of galaxy formation has been shaped by those galaxies that we have been able to observe. The study of these faint galaxies is key to refining our knowledge of the processes driving the evolution of galaxies in the $\Lambda$ Cold Dark Matter ($\Lambda$CDM) framework.

In 2015, \cite{vanDokkum:15} identified 47 remarkably large and faint LSBGs in the Coma cluster, with  effective radii $R_e>1.5$~kpc and central surface brightnesses $\mu_{0,g}>24$ mag arcsec$^{-2}$, popularising the term `ultra-diffuse galaxies' (UDGs). This sparked numerous studies of UDGs in the Coma cluster \citep[e.g.][]{Koda:15,Yagi:16} and in other clusters \citep[e.g.][]{Mihos:15,vanderBurg:16,mancerapina:19a,Lim:20,Iodice:20,LaMarca:22}. UDGs have also been found in groups and lower density environments  \citep[e.g.][]{Leisman:17,Spekkens:18,Janowiecki:19,mancerapina:20,Kadowaki:21}. UDGs in cluster environments tend to host older, redder stellar populations, while those in less dense environments have been found to have younger, bluer stellar populations and significant cold gas reservoirs \citep[e.g.][]{vanDokkum:15,Leisman:17}. 

Using the ALFALFA survey, \cite{Leisman:17} found 30 H\,{\sc i}-bearing UDGs (HUDs) that follow the definition from \cite{vanDokkum:15}, and a broader sample of 115 HUDs following the definition from \cite{vanderBurg:16}. The UDG selection criteria are somewhat arbitrary. There are multiple definitions based on different surface brightness limits and size cuts \citep{vanDokkum:15,Koda:15,vanderBurg:16,Forbes:20}, orientation considerations \citep{He:19} as well as scaling relations \citep{Lim:20}. Using the Romulus simulations, \cite{vanNest:22} investigate 5 different UDG definitions, finding that the number of sources identified as UDGs changes by up to $\sim45$~per~cent. They also demonstrate that the classification is highly dependent on the viewing angle, with this effect being strongest for UDGs in low-density environments. An additional source of difficulty in identifying UDGs arises from the distance requirement needed for the selection criteria. The Systematically Measuring Ultra-diffuse Galaxies \citep[SMUDGes;][]{Zaritsky:19,Zaritsky:23} presents 7070 UDG candidates, however without distance information they are unable to confirm all of the candidates.

%
Not only is there a varied population of UDGs across different environments, the different definitions in the literature affect the sample selection, adding to the diversity of these extreme sources. Consequently, a range of formation mechanisms and evolutionary pathways are required to explain them. UDGs with high abundances of globular clusters have been suggested to be `failed galaxies' in {massive dark matter haloes  \citep[$\sim 10^{11} - 10^{12}$ M$_{\odot}$; ][]{vanDokkum:16,Forbes:20}. } {However, accurately determining the number of globular clusters associated with a UDG is challenging, and claims of high globular cluster abundances in some such systems remain subject to debate \citep[e.g.][]{Saifollahi:21,Saifollahi:22}.} Additionally, UDGs have also been shown to form in regular dwarf sized haloes through both internal and external processes. Internal factors include high spin haloes, which can cause inefficient star formation {and large sizes} \citep{Amorisco:16,Rong:17,mancerapina:20,Benavides:24}, and  supernovae driven gas outflows, which can lead to an extended stellar distribution \citep{DiCintio:17}. On the other hand,  dense environments, like clusters and groups, can cause galaxies to `puff up' though tidal heating while ram pressure stripping removes gas and quenches star formation { \citep{Jiang:19,Sales:20,Tremmel:20,Junais:22}.}  In isolated environments, early mergers can trigger a temporary boost in spin and cause star formation to be redistributed to the outskirts of galaxies, which ultimately results in the formation of a UDG \citep{Wright:21}. UDGs can also form as quenched, gas-poor backsplash galaxies that were originally part of denser environments \citep{Benavides:21}. Furthermore, UDGs can form without dark matter, {or in very low density dark matter haloes \citep{mancerapina:19,Kong:22,Mancerapina:22,Afruni:25}}. HUDs can form as tidal dwarf galaxies such as the almost dark clouds in the Hickson Compact Group 16 \citep{roman:21} and the Klemola 13 group \citep{obeirne:24}. \cite{Watts:24} present the discovery of an UDG that formed at the end of an extended tidal tail associated with the past interaction of NGC 3785 with a gas-rich galaxy.  Red, gas-poor, dark matter free UDGs can form through high velocity galaxy collisions \citep{Silk:19,vanDokkum:22}.

Despite the thousands of observations of UDGs/UDG candidates, no confirmed UDG pairs have been identified to date. Across the 7070 SMUDGes UDG candidates, only 8 ongoing merger and 7 post merger low surface brightness systems were identified \citep{Zaritsky:25}. {Some gas-rich UDGs have been identified as potential post-merger systems or as having undergone interactions with gas-rich dwarf galaxies \citep[e.g.][]{Bennet:18,Fielder:23, Buzzo:25}.} 

{Using} neutral atomic hydrogen (H\,{\sc i}) data from the Widefield ASKAP L-band Legacy All-sky Survey \citep[WALLABY;][]{Koribalski:20}, we have identified a pair of HUDs in the Hydra field of the pilot survey (source name WALLABY J104513-262755), both of which satisfy the \cite{vanDokkum:15} UDG selection criteria. WALLABY is useful for studying interacting galaxies, as demonstrated by \cite{Huang:24}, who study the star formation in gas-rich galaxy pairs. They find that low stellar mass galaxies may initially experience suppressed star formation following the halo encounter, before undergoing enhanced star formation as their H{\sc i} discs overlap. Furthermore, WALLABY has been shown to be a powerful probe of the low surface brightness Universe, with 315 LSBGs identified in just three of the pilot fields (making up almost 20 per cent of the sample), 75 per cent of which had not been previously catalogued \citep{obeirne:25}. Additionally, using the pre-pilot WALLABY data, \cite{For:23} analysed UDG candidates identified in SMUDGes, finding one H\,{\sc i} bearing low surface brightness dwarf and six putative UDGs without H\,{\sc i} in the Eridanus supergroup. 

\cite{obeirne:25} found that the LSBGs had diverse properties, with stellar masses spanning four orders of magnitude from $\sim 10^6$ to $\sim 10^{10}$ M$_{\odot}$. Here, we study the galaxies that represent the extreme tail of this population, in both size and surface brightness. HUDs are important galaxies for investigating factors that cause star formation to be suppressed without the removal of gas.  

This paper is structured as follows. Sections \ref{sec:data} and \ref{sec:photometry} outline the data and methods used in this study. Section \ref{sec:sample} presents the WALLABY UDG sample. Section \ref{sec:pair} showcases the UDG pair, including its photometric properties, H{\sc i} properties and environment. Finally, Section \ref{sec:comparison} compares the pair with other HUDs and Section \ref{sec:concl} summarises our results. Throughout the paper we {we adopt the AB magnitude system and} use velocity in the optical convention .


\section{Data}
\label{sec:data}

The data used in this study is outlined in detail in \cite{obeirne:25}. To summarise, we use H\,{\sc i} data from fields in the WALLABY pilot survey phase 1 \citep[the Hydra and NGC 4636 fields,][]{Westmeier:22}, and phase 2 \citep[the NGC 4808 and NGC 5044 fields,][]{Murugeshan:24}. {Additionally, we use the currently available full survey data that has deep optical imaging available (the WALLABY 0051-53,  0117-59,  0125-53,  0150-16,  0154-21,  0156-59,  0237-37,  0241-32,  0325-21,  0931-21,  1451-21,  1506-32,  1017-21 and  2051-53 fields). Across the pilot and full surveys used in this study, there are a total of 3532 H{\sc i} detections. } WALLABY has a 30 arcsec beam and channel resolution of 4~km~s$^{-1}$.  WALLABY has a target noise level of 1.6 mJy beam$^{-1}$ per channel, which corresponds to a $5\sigma$ column density of $8.6\times10^{19}$cm$^{-2}$ across 20 km s$^{-1}$. The Hydra field has a slightly higher noise level of 1.85 mJy beam$^{-1}$ per channel. The phase 2 data {and currently available full survey data} have an observed median rms noise of 1.7 mJy beam$^{-1}$ per channel, which corresponds to a 5$\sigma$ H\,{\sc i} column density sensitivity of $ \sim 9.1 \times 10^{19}$ cm$^{-2}$ across 20 km~s$^{-1}$. The sources in WALLABY are identified using version 2 of the Source Finding Application \citep[SoFiA;][]{Serra:15,Westmeier:21}. The optical imaging used in this study is from the Legacy Survey DR10, which has seeing on the order of 1 arcsec, and has $3\sigma$ limiting surface brightnesses of 29.8, 29.4, 27.7 and 28.0 mag arcsec$^{-2}$ in the $g$, $r$, $i$ and $z$-bands respectively, (measured in $10\times10$ arcsec boxes as measured by \citealt{obeirne:24} following the depth definition by \citealt{Roman:20}). We use ultraviolet imaging from the Galaxy Evolution Explorer \citep[GALEX;][]{Martin:05}. GALEX provides imaging in the near ultraviolet (NUV; 1770-2730 Å) and the far ultraviolet (FUV; 1350-1780 Å). Finally, infrared imaging from Wide-field Infrared Survey Explorer \citep[WISE;][]{Wright:10} is used in this study. WISE provides data over the entire sky in four mid-infrared bands: W1 (3.4 $\mu$m), W2 (4.6 $\mu$m), W3 (12 $\mu$m) and W4 (22 $\mu$m).

\section{Methods}
\label{sec:photometry}

\subsection{Photometry}

To measure the optical properties and identify the UDGs, we follow the methods of \cite{obeirne:25} to fit Sérsic models to the Legacy Survey images, and calculate the stellar mass.  We summarise the method here. We use the python package \textsc{AstroPhot} \citep{Stone:23} to model the $g$ and $r$-bands together (or $g$ and $i$-bands if the $r$-band is not available) along with any bright foreground stars. This allows us to optimise the signal-to-noise and get consistent photometry across the bands, and consequently obtain good colours. Additionally, the models take the point spread function (PSF) into account. To create the PSF, we fit a Moffat PSF profile to five stars in each band.  \textsc{AstroPhot} fits seven parameters in the Sérsic model: the centre coordinates, position angle ($PA$), axis ratio ($q$), the brightness at the half-light radius ($I_{e}$), the effective radius ($R_{e}$) and the Sérsic index ($n$). In our joint model, all parameters are fit together for both bands, except $I_{e}$ and $n$, which are allowed to vary. In addition to the model parameters, the models allow us to obtain the central surface brightness and total flux of the UDGs.  We account for Galactic extinction using the correction method outlined in \cite{Yuan:13}, adopting the $R(a)$ values from \cite{Schlegel:98} and the $E(B-V)$ values from \cite{Schlafly:11} (the mean value provided within a 5 arcmin radius of the galaxy coordinates).  Additionally, we apply $k$-corrections following \cite{Chilingarian:10}.

To calculate the stellar mass, we use the relation between stellar mass to light ratio ($\frac{\Upsilon^*}{\rm M_{\odot}/L_{\odot}}$) and $g-r$ colour from \cite{Du:20}, which is calibrated specifically for LSBGs:
\begin{equation}
    \log_{10}(\frac{\Upsilon^*}{\rm M_{\odot}/L_{\odot}}) = a + b({\rm colour}) ,
    \label{Mstar_lsb}
\end{equation}
where $a=-0.857$ and $b=1.558$ are the coeffients if $g-r$ is used, and $a=-1.152$ and $b = 1.328$ if $g-i$ is used. {This relation assumes the initial mass function of \cite{Chabrier:03}. } The uncertainty in the stellar mass, $\Delta \log( \frac{M_*}{\rm M_{\odot}}) = 0.12$, is taken to be the scatter in the $\log_{10}(\Upsilon^{*})$ relation from \cite{Du:20}. 


As the sources in the GALEX images have low signal to noise ratios, instead of fitting Sérsic models, we measure the total flux inside the {elliptical} optical apertures with radii$=2\times R_e$ {and the measured axis ratios}. {For one edge-on UDG in the sample (WALLABY J104513-262755-UDG-2), the original {elliptical} aperture is smaller than the FWHM angular resolution of the GALEX image. As a result, a circular aperture is used for this source instead.} The SFR is calculated solely from the UV using the relation from \cite{Hao:11}:
\begin{equation}
    \log_{10}\left(\frac{\rm SFR}{{\rm M}_{\odot}~{\rm yr}^{-1}}\right) = \log_{10}\left(\frac{\rm L_{\rm FUV}}{\rm erg~s^{-1}}\right) - 43.35
    \label{eq:SFR_GALEX}
\end{equation}
where $L_{\rm FUV}$ is the reddening- and attenuation-corrected FUV luminosity. However, if $FUV-NUV\le0$, the internal dust attenuation correction can not be applied, as the $A_{\rm FUV}$ correction parameter from Equation 16 in  \cite{Hao:11} becomes unphysical. {This relation assumes the initial mass function of \cite{Kroupa:03}. }Additionally, we have also applied the $k$-corrections following \cite{Chilingarian:10}. The uncertainty in the SFR is { $\pm0.12$} dex (from the scatter of the relation in Equation \ref{eq:SFR_GALEX}). One source (WALLABY J033402-205037) is located beyond the  coverage of GALEX. As was the case with the low surface brightness sample studied {by} \cite{obeirne:25}, the UDGs appear to be dust poor, lacking reliable detections in the WISE W3 and W4 bands.


\subsection{H I Properties and Kinematics}
\label{sec:HIprop}





To study the H{\sc i} properties and kinematics of the UDGs in our sample we take advantage of the WALLABY catalogues and data products. This includes the central frequency, {the width of the H{\sc i} spectral line at $50\%$ of the peak flux density ($w_{50}$)}, total integrated flux and associated error, and the masked moment 0 and moment 1 maps. We apply the WALLABY statistical flux corrections to account for the flux deficit of the faint WALLABY sources compared to the flux that would be recovered by single dish observations. The phase 1 corrections are applied to sources from the Hydra and NGC 4636 fields \citep{Westmeier:22} and the phase 2 corrections are applied to those from the NGC 4808, NGC 5044 and full survey fields \citep{Murugeshan:24}. We calculate the H{\sc i} mass from the total integrated flux \citep{Meyer:17}, propagating the flux uncertainty to calculate the uncertainty in the H{\sc i} mass. {Because UDG-1 and UDG-2 are identified as a single source, WALLABY J104513–262755, we do not use the SoFiA emission line widths as done for the rest of the sample. Instead, we measure the $w_{50}$ and $w_{20}$ directly from the spectrum of each source, as shown in Section~\ref{sec:pair_hi}.} The only UDG in our sample that was able to be modelled by the WALLABY Kinematic Analysis Proto-Pipeline \citep[WKAPP;][]{Deg:22,Murugeshan:24} is WALLABY J125956-192430. This source has a particularly low rotation velocity and is studied in detail in \cite{Dudley:25}. {The other UDGs lack sufficient spatial resolution for reliable kinematic modelling. }

\section{The UDG sample}
\label{sec:sample}

We have identified a {core} sample of {10} UDGs that meet the criteria defined by \cite{vanDokkum:15} ($\mu_{g,0}\ge24$ mag~arcsec$^{-2}$ and $R_{e}\ge1.5$ kpc), {which we adopt as the sole UDG criteria for this work. For reference, we note that} {these UDGs also satisfy the \cite{vanderBurg:16} criteria, exhibiting $r$-band surface brightnesses at the effective radius fainter than 24 mag arcsec$^{-2}$, with the exception of one source for which r-band imaging is unavailable.} Additionally, for completeness we have included an additional {12} faint diffuse galaxies as part of a broader sample of UDGs. We do this because, as discussed in Section \ref{sec:intro}, the UDG criteria are ill-constrained, and using different model parameters and distances can affect the measured $\mu_{0,g}$ and $R_{e}$ values. {We note that all of the UDGs in the {core} sample continue to meet the \cite{vanDokkum:15} criteria even if their distances have been overestimated by 4 Mpc, (corresponding to a peculiar velocity of $ 300~\mathrm{km~s^{-1}}$ assuming $H_0 = 70~\mathrm{km~s^{-1}~Mpc^{-1}}$).} All UDGs in the broader sample have $\mu_{0,g}\ge23.7$ {mag arcsec$^{-2}$} and $R_e \ge 1.3$ kpc. {The combined core and broad WALLABY UDG sample exhibits high H{\sc i}-to-stellar mass ratios. The most extreme case is WALLABY J024013-393446, with a ratio of 27, while the sample as a whole has a mean ratio of 8.} 

WALLABY J104513-262755 is an H{\sc i} detection in the Hydra field of the Phase 1 WALLABY pilot survey. Although it was identified as a single H{\sc i} source by the automated SoFiA pipeline used by WALLABY, it hosts two distinct {H{\sc i} peaks which correspond to two optically-faint} galaxies that can be seen in the Legacy Survey images: {WALLABY~J104513-262755-UDG-1 and WALLABY~J104513-262755-UDG-2}. Hereafter, WALLABY~J104513-262755 refers to the pair, UDG-1 refers to the north-western galaxy and UDG-2 refers to the south-eastern galaxy. A full analysis on the pair, including how the individual H{\sc i} masses, central velocities and {$w_{20}$} values were measured, can be found in Section \ref{sec:pair}.

Table \ref{tab:sersic} presents the properties of the WALLABY UDGs we have identified. {The systemic velocities are given by $cz$, where $z$ is the redshift derived from the central frequency of the H {\sc i} emission line.} Two of the UDGs in our sample also feature in other WALLABY studies, and consequently for consistency, we use the parameters provided in those studies.  WALLABY J103508-283527 is a part of the interacting group Klemola 13 and was likely formed through tidal interactions \citep{obeirne:24}.  {WALLABY J125956-192430 is a relatively nearby and well resolved UDG at a distance of $7.3\pm0.3$ Mpc \citep[measured using the tip of the red giant branch;][]{Karachentsev:17} at a moderate inclination (58 deg).  \cite{Dudley:25} have performed detailed kinematic and mass modelling of this source, finding that it is dark matter dominated.} {The $w_{50}$ and H{\sc i} mass of WALLABY J032705-211110 are not listed in Table \ref{tab:sersic} because the H{\sc i} properties of this galaxy could not be disentangled from the shared H{\sc i} envelope with its companion galaxy (see Section \ref{sec:otherpair} for more details on this source).} Optical images with H{\sc i} contours overlaid and moment 1 maps of each UDG in the sample can be found in Appendix \ref{sec:appen_figs}.

\begin{table*}
\setlength{\tabcolsep}{2pt}
    \centering
    \caption{Photometric and H{\sc i} properties of the UDG sample. The horizontal line separates the {core} sample (top section) from the broader sample (bottom section). From left to right the coloumns are: the source name, central coordinates of the modelled source, {H{\sc i}} system velocity (cz), distance, {$w_{50}$} H{\sc i} emission line width, central surface brightness,  effective radius in kpc, minor to major axis ratio, colour, stellar mass, star formation rate and H{\sc i} mass. The distances provided are the luminosity distances (approximated by the Hubble law using the {heliocentric} velocities of the H{\sc i} detections), except for that of WALLABY J125956-192430, which has a distance measured using the tip of the red giant branch \citep{Karachentsev:17}. Colour values given in parentheses indicates the $g-i$ colour is given instead of the $g-r$ colour {because the $r$-band image was not available}. SFR values marked by $^*$ indicates that the internal dust extinction correction could not be applied. Blank SFR indicates that the source was not covered by GALEX. {J032705-211110-UDG refers to the UDG component of this WALLABY detection, rather than the companion galaxy. Only the optical properties are shown, as the H{\sc i} properties of this source cannot be disentangled from the shared H{\sc i} envelope of this system (see Section \ref{sec:otherpair}).} }
    \label{tab:sersic}
    \begin{tabular}{cccccccccccc}
    \hline
        &  Central Coordinates & $v_{sys}$ & $d$ & $w_{50}$ & $\mu_{0,g}$  & $R_e$ & $q$ & $g-r(i)$ &  $\log\left(\frac{M_{*}}{\rm M_{\odot}}\right)$ & $\log\left(\frac{SFR}{\rm M_{\odot}~yr^{-1}}\right)$& $\log\left(\frac{M_{HI}}{\rm M_{\odot}}\right)$  \\
        & J2000 & km s$^{-1}$& Mpc & km s$^{-1}$& {\scriptsize mag arcsec$^{-2}$} & kpc &  & mag & &  &    \\
        \hline
UDG-1 & 161.333$^{\circ}$,-26.488$^{\circ}$ & 4294 & 61.9 & 27.4 & 24.2 & 2.8$\pm$0.1 & 0.8$\pm$0.1 & 0.30 & 8.07$\pm0.12$ & -1.27$\pm0.12$ & 8.95$\pm$0.03 \\
UDG-2 & 161.346$^{\circ}$,-26.500$^{\circ}$ & 4260 & 61.9 & 37.5 & 24.6 & 2.2$\pm$0.1 & 0.2$\pm$0.1 & 0.22 & 7.16$\pm0.12$ & -2.10$^{*}\pm0.12$ & 8.60$\pm$0.04 \\
J103508-283427 & 158.785$^{\circ}$,-28.585$^{\circ}$ & 3286 & 47.4 & 47.0 & 26.3 & 6.1$\pm$0.2 & 0.6$\pm$0.0 & 0.50 & 8.00$\pm0.12$ & <-2.65 & 8.69$\pm$0.04 \\
J131403-120644 & 198.517$^{\circ}$,-12.113$^{\circ}$ & 6596 & 95.8 & 92.3 & 24.2 & 5.4$\pm$0.1 & 0.8$\pm$0.1 & (0.55) & 8.74$\pm0.12$ & -0.94$\pm0.12$ & 9.10$\pm$0.04 \\
J131132-144033 & 197.884$^{\circ}$,-14.676$^{\circ}$ & 2459 & 35.3 & 92.3 & 24.0 & 2.3$\pm$0.0 & 0.6$\pm$0.1 & 0.35 & 7.98$\pm0.12$ & -1.31$\pm0.12$ & 8.85$\pm$0.02 \\
 {J145109-214730} & 222.804$^{\circ}$,-21.805$^{\circ}$ & 3390 & 48.8 & 49.8 & 24.1 & 2.4$\pm$0.1 & 0.6$\pm$0.1 & 0.31 & 7.80$\pm0.12$ & <-3.60 & 8.73$\pm$0.07 \\
 {J144813-202638}  & 222.058$^{\circ}$,-20.443$^{\circ}$ & 4038 & 58.3 & 17.4 & 24.7 & 1.9$\pm$0.1 & 0.8$\pm$0.1 & 0.37 & 7.84$\pm0.12$ & <-2.67 & 8.53$\pm$0.04 \\
 {J033402-205037} & 53.508$^{\circ}$,-20.845$^{\circ}$ & 1758 & 25.2 & 17.8 & 24.5 & 1.9$\pm$0.0 & 0.6$\pm$0.0 & 0.29 & 7.35$\pm0.12$ &   & 7.96$\pm$0.06 \\
 {J205043-531619} & 312.678$^{\circ}$,-53.274$^{\circ}$ & 6967 & 101.3 & 60.8 & 24.4 & 4.2$\pm$0.1 & 0.6$\pm$0.0 & 0.37 & 8.24$\pm0.12$ & -1.35$\pm0.12$ & 9.24$\pm$0.05 \\
 {J032705-211110-UDG} & 51.780$^{\circ}$,-21.184$^{\circ}$ & 10697 & 157.0 &   & 25.0 & 5.9$\pm$0.2 & 0.4$\pm$0.1 & 0.59 & 8.58$\pm0.12$ & <-2.99 &   \\
 \hline
J125956-192430 & 194.983$^{\circ}$,-19.408$^{\circ}$ & 828 & 7.3 & 38.1 & 24.3 & 1.3$\pm$0.1 & 0.5$\pm$0.1 & 0.25 & 6.80$\pm0.20$ & -2.92$\pm0.12$ & 7.73$\pm$0.08 \\
J125311+032639 & 193.296$^{\circ}$,3.442$^{\circ}$ & 2794 & 40.2 & 16.9 & 23.7 & 3.1$\pm$0.1 & 0.7$\pm$0.1 & 0.32 & 7.98$\pm0.12$ & -1.31$\pm0.12$ & 8.71$\pm$0.04 \\
J125603+045202 & 194.014$^{\circ}$,4.867$^{\circ}$ & 1374 & 19.7 & 26.5 & 24.5 & 1.4$\pm$0.0 & 0.6$\pm$0.1 & 0.38 & 7.48$\pm0.12$ & <-3.01 & 7.77$\pm$0.04 \\
J131845-153954 & 199.690$^{\circ}$,-15.665$^{\circ}$ & 6482 & 94.1 & 24.6 & 23.8 & 7.9$\pm$0.2 & 0.7$\pm$0.1 & (0.36) & 8.37$\pm0.12$ & -0.16$\pm0.12$ & 9.42$\pm$0.03 \\
J132030-202723 & 200.128$^{\circ}$,-20.457$^{\circ}$ & 1626 & 23.3 & 21.4 & 24.2 & 1.3$\pm$0.0 & 0.7$\pm$0.1 & 0.32 & 7.47$\pm0.12$ & <-3.23 & 7.58$\pm$0.04 \\
J133424-141834 & 203.601$^{\circ}$,-14.311$^{\circ}$ & 6644 & 96.5 & 28.6 & 23.7 & 3.5$\pm$0.1 & 0.8$\pm$0.1 & 0.33 & 8.39$\pm0.12$ & -1.43$^{*}\pm0.12$ & 9.15$\pm$0.05 \\
J134247-193454 & 205.696$^{\circ}$,-19.582$^{\circ}$ & 1411 & 20.2 & 36.7 & 24.2 & 1.4$\pm$0.0 & 0.8$\pm$0.0 & 0.11 & 7.32$\pm0.12$ & <-2.97 & 8.49$\pm$0.02 \\
 {J024013-393446} & 40.055$^{\circ}$,-39.579$^{\circ}$ & 1954 & 28.1 & 36.0 & 23.7 & 1.6$\pm$0.0 & 0.5$\pm$0.0 & 0.27 & 7.30$\pm0.12$ & -2.22$\pm0.12$ & 8.42$\pm$0.03 \\
J123308-003203 & 188.283$^{\circ}$,-0.533$^{\circ}$ & 724 & 10.4 & 19.7 & 23.7 & 1.8$\pm$0.0 & 0.8$\pm$0.0 & 0.44 & 7.71$\pm0.12$ & -1.68$\pm0.12$ & 7.74$\pm$0.02 \\
 {J010649-601631} & 16.708$^{\circ}$,-60.276$^{\circ}$ & 4934 & 71.4 & 84.1 & 23.9 & 3.0$\pm$0.1 & 0.5$\pm$0.0 & 0.26 & 8.03$\pm0.12$ & -1.29$\pm0.12$ & 9.10$\pm$0.05 \\
 {J093302-233940} & 143.270$^{\circ}$,-23.668$^{\circ}$ & 9818 & 143.8 & 70.5 & 23.7 & 4.3$\pm$0.1 & 0.8$\pm$0.1 & 0.39 & 8.86$\pm0.12$ & -1.08$^{*}\pm0.12$ & 9.28$\pm$0.06 \\
 {J102132-213357} & 155.396$^{\circ}$,-21.575$^{\circ}$ & 3177 & 45.8 & 65.5 & 23.8 & 1.6$\pm$0.0 & 0.8$\pm$0.1 & 0.29 & 7.75$\pm0.12$ & -1.87$\pm0.12$ & 8.47$\pm$0.06 \\

\hline
    \end{tabular}
\end{table*}

\subsection{SMUDGes}
\label{sec:smudges}

{The NGC 4636 field, the NGC 4808 field and the WALLABY full survey fields used in this study overlap with the area covered by the SMUDGes catalogue. SMUDGes uses an automated search for UDGs in the Legacy Survey DR9 images using a modified version of a deep learning model followed by visual confirmation \citep[for details on the method, see][]{Zaritsky:19,Zaritsky:21,Zaritsky:23}. We only consider SMUDGes candidates that were classified as `good' candidates by their visual inspection. The UDG candidate selection is based on the criteria $\mu_{0,g} \ge 24$ mag arcsec$^{-2}$ and $R_e \ge 5.3$ arcsec. Due to the lack of distance measurements, these candidates cannot be confirmed as UDGs, since the UDG definition requires $R_e$ in physical units rather than angular size. H{\sc i} observations can provide this information. }

Of the UDGs in our WALLABY sample, six are also present in SMUDGes: three from the core sample { (WALLABY J205043-531619, WALLABY J032705-211110 and WALLABY J033402-205037) and three from the broad sample (WALLABY J125311+032639, WALLABY J125603+045202, and WALLABY J010649-601631).} Additionally, there are four other SMUDGes candidates that are detected in the WALLABY data, however, we do not classify them as UDGs {  (WALLABY J033408-232125, WALLABY J125604+034843, WALLABY J124232-012111 and WALLABY J125313+042746).} We also identify four UDGs that are within the SMUDGes footprint that are not listed in their catalogue (WALLABY J123308-003203, WALLABY J145109-214730, WALLABY J144813-202638, and WALLABY J024013-393446), likely due to our use of the deeper DR10 images and the advantage of having H{\sc i} positions from WALLABY. We do not conduct a separate H{\sc i} search at the coordinates of SMUDGes UDGs in the unmasked H{\sc i} data. {Table \ref{tab:smudges} in Appendix \ref{sec:appen_smudges} provides the SMUDGes names associated with the corresponding WALLABY sources. }


\section{The UDG pair}
\label{sec:pair}

\begin{figure}
    \centering
    \includegraphics[width=\linewidth]{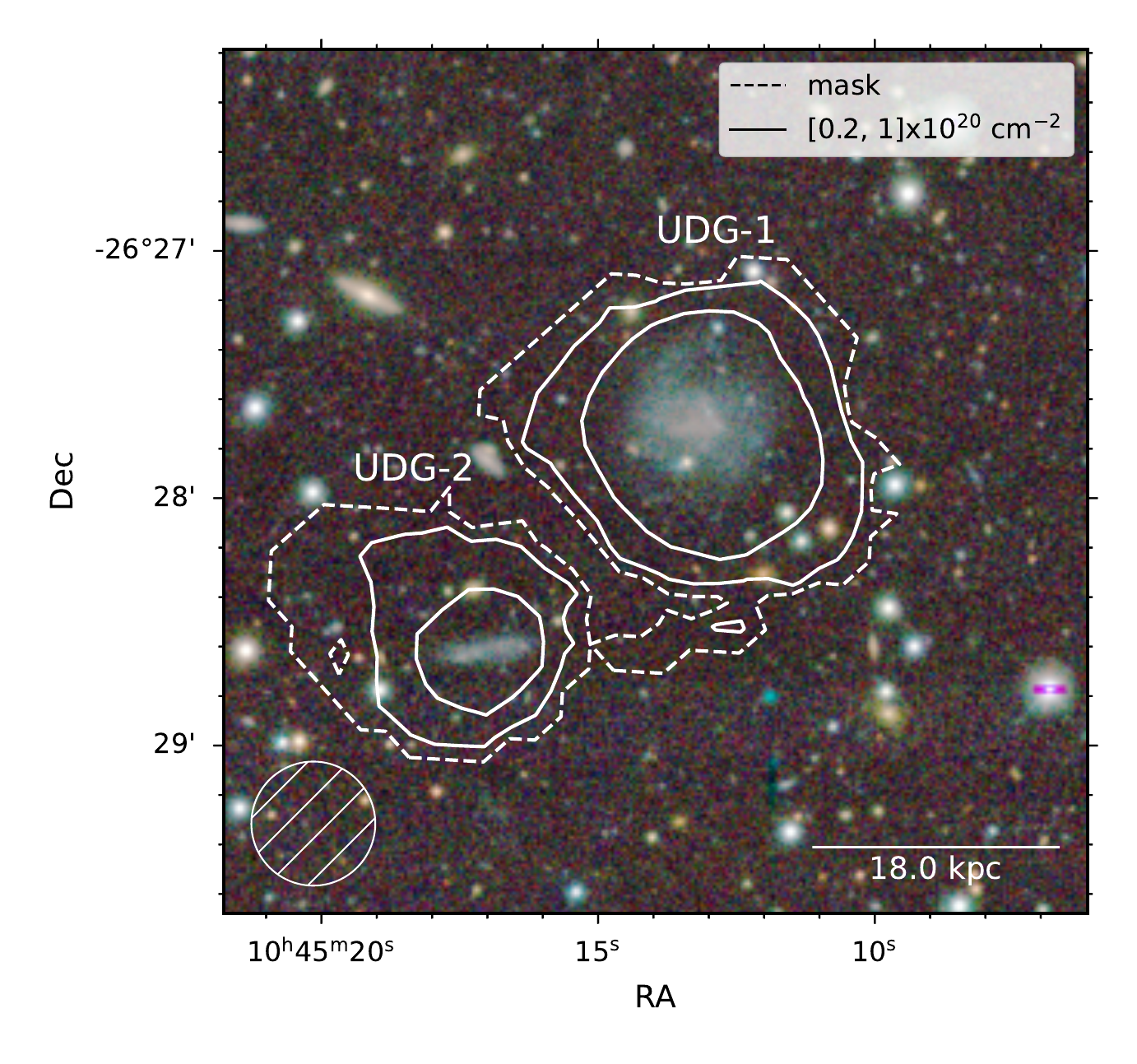}
    \caption{The three colour ($grz$-band) Legacy Survey image of WALLABY J104513-262755.  The H{\sc i} contours levels equal to the local column density rms ($2\times10^{19}$ cm$^{-2}$) and $1\times10^{20}$ cm$^{-2}$ are overlaid in white. The dashed contour represents the edge of the SoFiA mask. The north-western galaxy is labelled UDG-1 and the south-eastern galaxy is labelled UDG-2. The WALLABY beam (30 arcsec) is shown in the lower left corner.}
    \label{fig:legacy_HIcontours}
\end{figure}

WALLABY J104513-262755 is the first detection of a {UDG-UDG} pair, and these two galaxies had not been previously catalogued in the NASA/IPAC Extragalactic Database (NED) prior to WALLABY. Figure \ref{fig:legacy_HIcontours} shows the three-colour image constructed from the $g$, $r$ and $z$-band Legacy Survey DR10 images with H{\sc i} contours overlaid. There is evidence to suggest that H{\sc i}-bearing UDGs may simply lie at the extreme end of the gas-rich LSBG regime, rather than forming a distinct population \citep[e.g.][]{Motiwala:25}. There are varied definitions of UDGs, as discussed in Section \ref{sec:intro}, and the status of these galaxies as UDGs is dependent on the distance used. Regardless, this pair of low mass, low surface brightness, diffuse galaxies with {H{\sc i}} gas near the Hydra cluster is a unique system which can give us insight into the evolution of galaxies in this regime. In this work we use Hydra cluster parameters consistent with previous WALLABY studies \citep{Wang:21, obeirne:24}. We adopt a distance of $47.5 \pm 1.4$ Mpc and a heliocentric velocity of 3686 km s$^{-1}$ for the Hydra cluster \citep{Kourkchi:17}. The cluster has a virial radius of $r_{200} \sim 1.35$ Mpc, and the mass enclosed within this radius is $M_{200} \sim 3.02\times10^{14}$ M$_{\odot}$ {\citep{Reiprich:02}}. The  velocity dispersion, $\sigma = 620$ km s$^{-1}$, is estimated from $M_{200}$ using the scaling relation from \citep{Evrard:08}.  {We note that} there is some tension in the literature whether the Hydra cluster is in fact virialised \citep[e.g.][]{Hayakawa:04} or not \citep[e.g.][]{Fitchett:88,Lima-Dias:21}. Throughout this section we will discuss {the photometric properties of the UDG-UDG pair, their H{\sc i} properties,} and their relationship with their local and large-scale environment. 

\subsection{Photometric properties}
\label{sec:pair_phot}

\begin{figure*}
     \centering
       \begin{subfigure}[t]{0.46\textwidth}
         \centering
         \includegraphics[width=\textwidth]{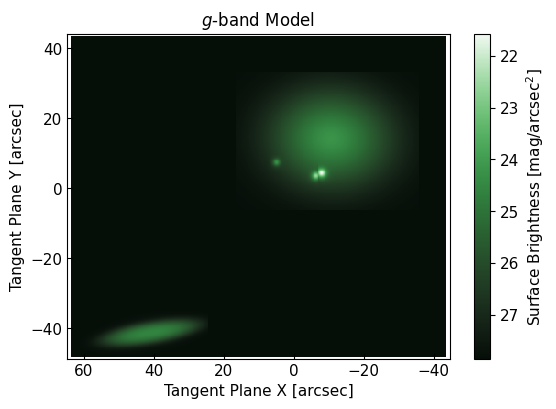}
         \caption{ }
         \label{fig:model}
     \end{subfigure}
     \begin{subfigure}[t]{0.5\textwidth}
         \centering
         \includegraphics[width=\textwidth]{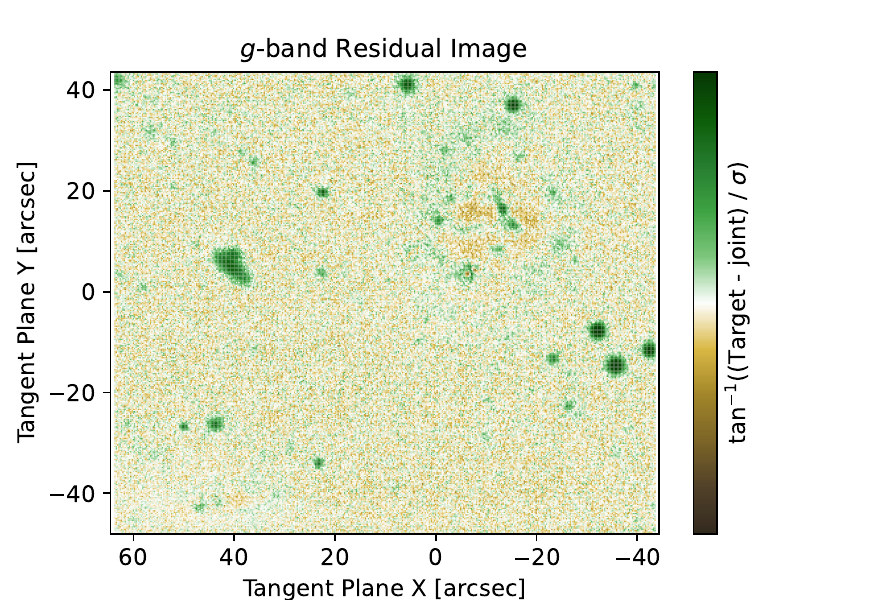}
         \caption{ }
         \label{fig:res}
     \end{subfigure}
     \begin{subfigure}[t]{0.64\textwidth}
         \centering
         \includegraphics[width=\textwidth]{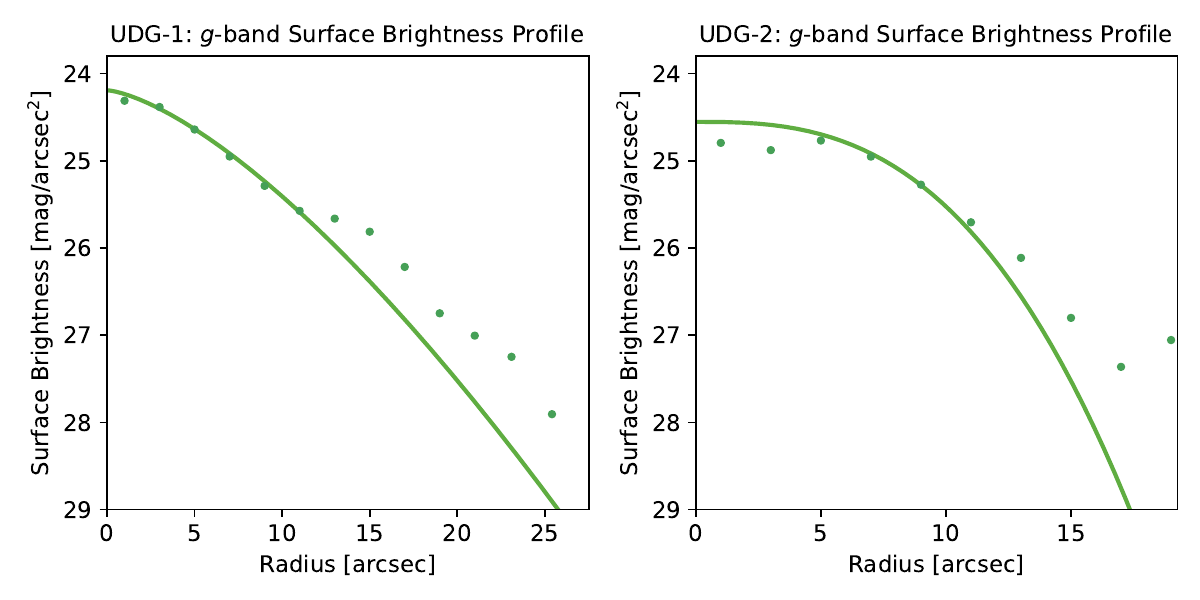}
         \caption{ }
         \label{fig:radial}
     \end{subfigure}
     \hfill
     \begin{subfigure}[t]{0.345\textwidth}
         \centering
         \includegraphics[width=\textwidth]{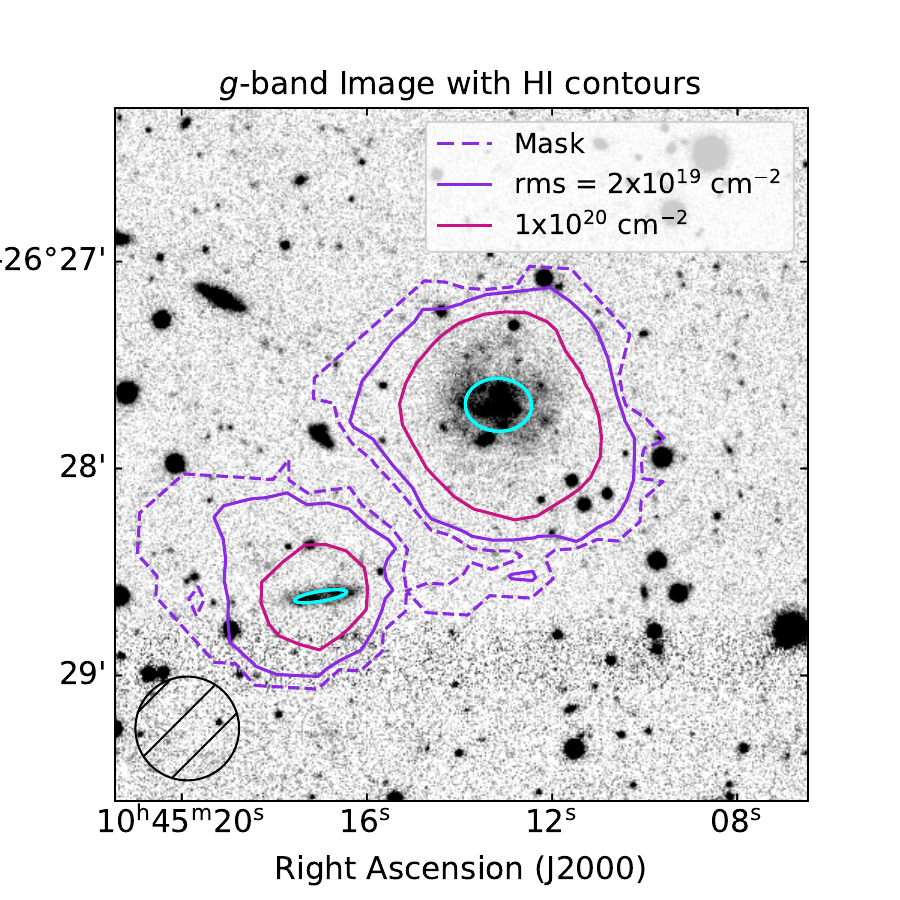}
         \caption{ }
         \label{fig:overlay}
     \end{subfigure}

    \caption{ Sérsic models of UDG-1 and UDG-2 in the $g$-band. (a) Model image of UDG-1, UDG-2 and three foreground stars. (b) Residual image. (c) Radial {$g$-band} surface brightness profiles of UDG-1 (left) and UDG-2 (right). The data profile, shown by the points, is the median of pixel values at a given radius. The model profile is shown by the solid line. {The Sérsic indices of UDG-1 and UDG-2 are 0.7 and 0.4 respectively.} (d) $g$-band image with H\,{\sc i} contours {(purple and pink) and optical aperture (cyan) overlaid. The dashed line corresponds to the edge of the SoFiA mask, and the lowest contour level shown represents the local rms. The optical apertures have the central coordinates, position angle and axis ratio from the Sérsic models and a radius equal to $R_e$.} The WALLABY beam is shown in the lower left corner.}
    \label{fig:gband}
\end{figure*}

Figure \ref{fig:gband} shows the UDG pair model image, residual image, the $g$-band radial surface brightness profiles, and the optical image with H\,{\sc i} contours and optical apertures (with radius = $R_e$, the PA, $q$ and central coordinates from the model) overlaid.   We note that slight variations in the central coordinates of the galaxies lead to modest changes in the models, however all resulting models consistently meet the \cite{vanDokkum:15} UDG classification criteria. {Additionally, Figure \ref{fig:galex} in Appendix \ref{sec:append_models} } shows the NUV and FUV GALEX images with H\,{\sc i} contours and optical apertures overlaid (with radius = $2R_e$, note {that} a circular aperture is used here for UDG-2 as discussed in Section \ref{sec:photometry}).  Figure \ref{fig:wise}  shows the WISE images with H\,{\sc i} contours and the optical apertures from the Sérsic models overlaid, however, UDG-1 is only {just visible} in the W1 and W2 bands, and UDG-2 is not detected at all in any of the four WISE bands.

In Figure \ref{fig:legacy_HIcontours}, {at the centre of UDG-1, there is a horizontally orientated structure that appears redder than the surrounding stellar light. This feature may indicate the presence of a bar.} While bar structures in UDGs are rare, they are not unprecedented. For instance, UDG8, studied as part of the Looking into the faintEst WIth MUSE (LEWIS) project on Hydra cluster UDGs, also exhibits signatures indicative of a bar \citep{Buttitta:25}. Low mass galaxies are typically stable against bar formation, however {bars} can be induced through tidal interactions \citep{Gajda:18,Lazar:24}.


The small optical axis ratio of UDG-2 ($q=0.2$) suggests that if the galaxy has a stellar disc, we must be viewing it edge-on. In this case it is remarkable that UDG-2 still meets the \cite{vanDokkum:15} UDG criteria without an inclination correction. If this galaxy were to be observed face-on, its central $g$-band surface brightness would measure just {26.3 mag arcsec$^{-2}$ instead of 24.6 mag arcsec$^{-2}$. We account for the effect of inclination using the expression $\mu_{0,face} = \mu_{0,edge} - 2.5\log(q)$ where $\mu_{0,face}$ and $\mu_{0,edge}$ are the edge-on and face-on central surface brightnesses, and $q$ is the minor-to-major axis ratio from the Sérsic fit.} This central surface brightness is the same as that of WALLABY J103508-283427. WALLABY J103508-283427 was initially identified as a dark cloud \citep{obeirne:24}, with no optical counterpart visible in the Legacy Survey DR9, and an extremely faint optical counterpart just visible in DR10. Thus if UDG-2 were to be viewed face-on rather than edge-on,  it would likely also lie on the threshold of being considered a `dark' source. This highlights that perhaps some dark sources, such as those  identified in \cite{obeirne:25}, could be face-on LSBGs that would have been optically visible had they been viewed from a different orientation.


Considering all the multiwavelength data adds nuance to the interpretation. The non-detection in the WISE (W3 or W4) imaging {may imply} a lack of dust (i.e. UDG-2 is optically thin). A closer look at the radial surface brightness profile in Figure \ref{fig:radial} reveals that the flat central region is inconsistent with an edge-on {circularly-symmetric} disc. In this case, the central surface brightness profile should appear to be more peaked, as the line of sight passes through more and more of the disc towards the centre. Furthermore, there is a $\sim 35^{\circ}$ offset between the optical position angle and the direction of the gradient in the velocity field (see Figure \ref{fig:mom1}; the H{\sc i} properties are discussed in the next subsection). This implies that, if UDG-2 is an edge-on disc, the H{\sc i} gas is strongly disturbed compared to the stars. This could be the result of interactions with UDG-1 (see Section \ref{sec:int} for this discussion).  Alternatively, rather than an edge-on disc, UDG-2 could be interpreted as a lower inclination disc that is traced by H{\sc i} gas and not the optical morphology. The optical counterpart may reflect the stochastic star formation, which is common in low-mass galaxies \citep[e.g.][]{Emami:19,Hopkins:23}. Additionally, we may only be viewing the brightest central region, while the extended disc lies below the sensitivity of the Legacy Survey, similar to the gas-rich UDG AGC 114905 \citep{mancerapina:24}, where the position angle from the shallower optical data suggests a different angle from the H{\sc i} kinematics, while the deeper optical data reveals the extended disc with a position angle that matches the kinematics. 

\begin{figure}
    \centering
    \includegraphics[width=0.95\linewidth]{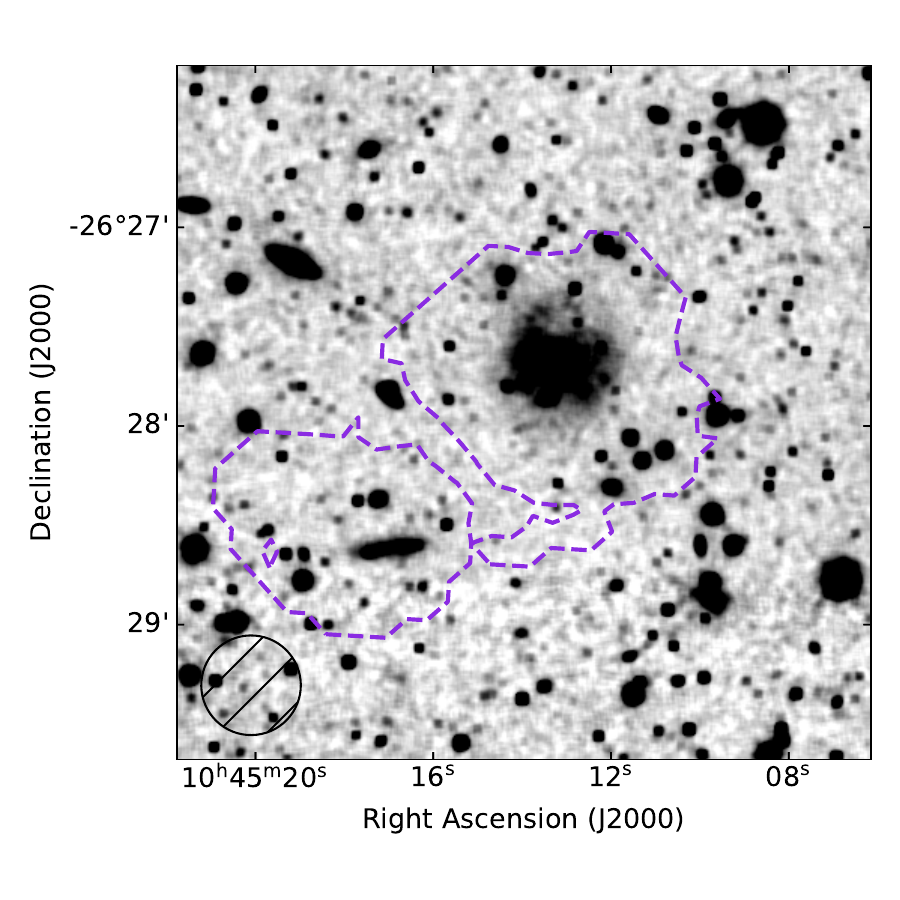}
    \caption{Smoothed, co-added image created with the $g$, $r$, $i$ and $z$-band images and by convolving with a 2D boxcar kernel with a size of 2.6 arcsec by 2.6 arcsec. The dashed contour represents the edge of the SoFiA mask.}
    \label{fig:coadd}
\end{figure}

To search for potential faint, diffuse structure in the stellar component of the galaxy, we create a smoothed, co-added image of the pair shown in Figure \ref{fig:coadd}. To create this image, we stack the $g$, $r$, $i$ and $z$-band images and convolve it with a 2D boxcar kernel with a size of 2.6 arcsec by 2.6 arcsec. {While} we do not see any signatures of faint tidal tails, {this does not rule out that the pair are interacting. Using the \textsc{New}\textsc{Horizon} cosmological simulation, \cite{Martin:22} create mock images to investigate the visibility of tidal features in the upcoming LSST. They find that an $r$-band $3\sigma$ limiting surface brightness (measured in $10\times10$ arcsec box) of 32 mag arcsec$^{-2}$ is required to detect $\gtrsim 25$ per cent of the area and $\gtrsim 80$ per cent of the total flux of tidal features for Milky Way or greater mass galaxies. At a limiting surface brightness of 29.5 mag arcsec$^{-2}$ (comparable with the Legacy Survey images), the total flux from tidal features is expected to drop to $\gtrsim 60$ per cent. They highlight that these values are expected to be lower in lower-mass galaxies, such as our UDG pair. The tidal debris from the UDG pair may remain invisible in the optical, }
{however, the H{\sc i} SoFiA mask does suggest the presence of some tidal features.}


\subsection{H I properties}
\label{sec:pair_hi}

\begin{figure*}
     \centering
    \begin{subfigure}[t]{0.48\textwidth}
         \centering
         \includegraphics[width=\textwidth]{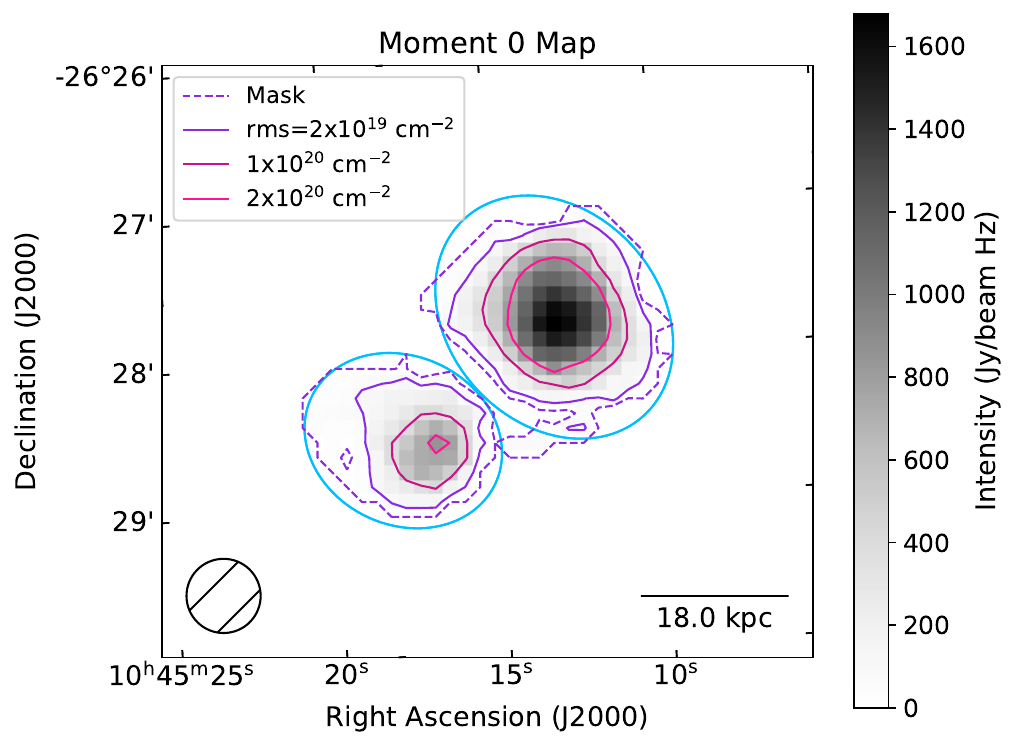}
         \caption{ }
         \label{fig:mom0}
     \end{subfigure}
     \hfill
    \begin{subfigure}[t]{0.48\textwidth}
         \centering
         \includegraphics[width=\textwidth]{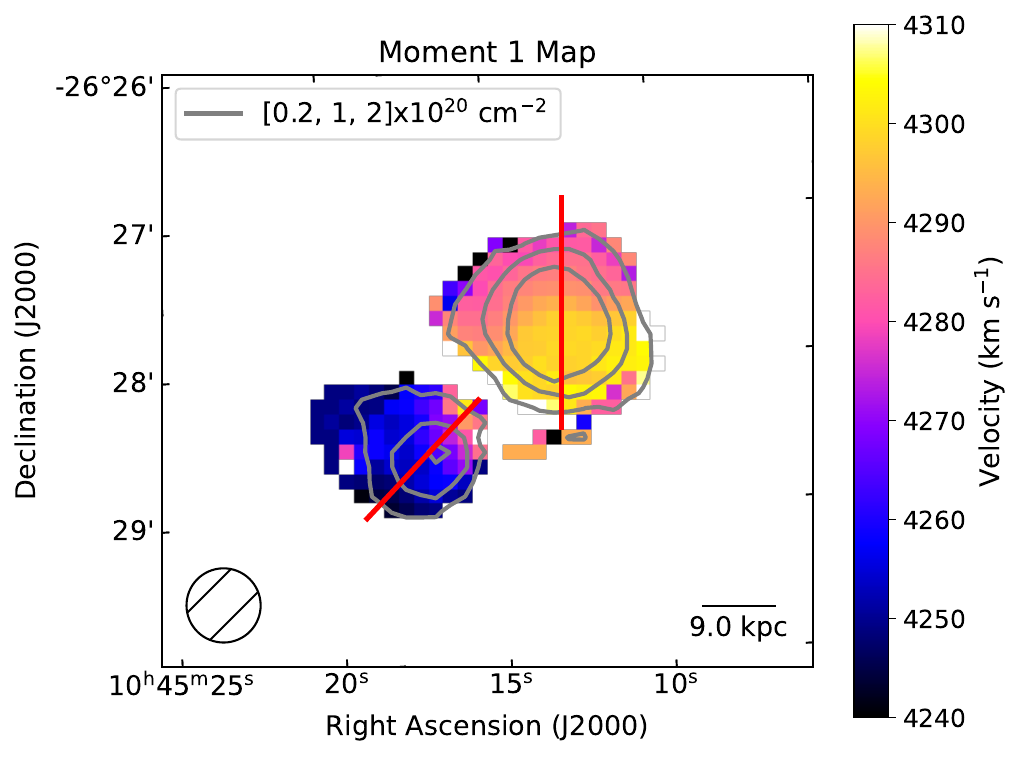}
         \caption{}
         \label{fig:mom1}
     \end{subfigure}

     \begin{subfigure}[t]{0.48\textwidth}
         \centering
         \includegraphics[width=\textwidth]{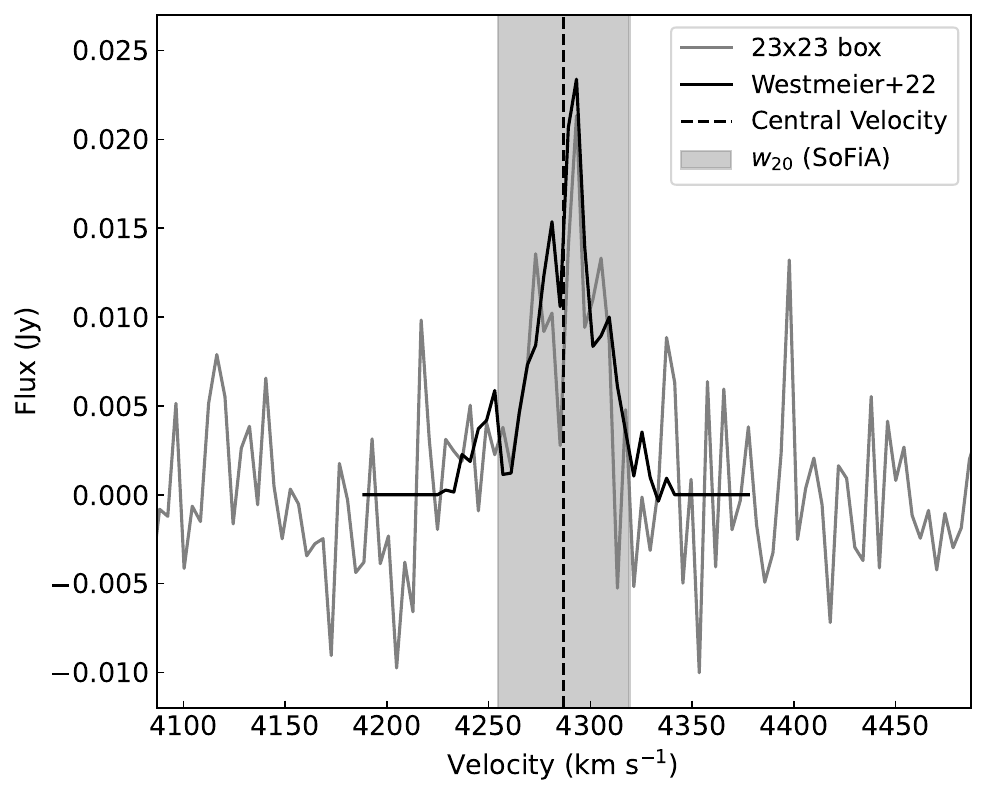}
         \caption{ }
         \label{fig:spec}
     \end{subfigure}
     \hfill
     \begin{subfigure}[t]{0.48\textwidth}
         \centering
         \includegraphics[width=\textwidth]{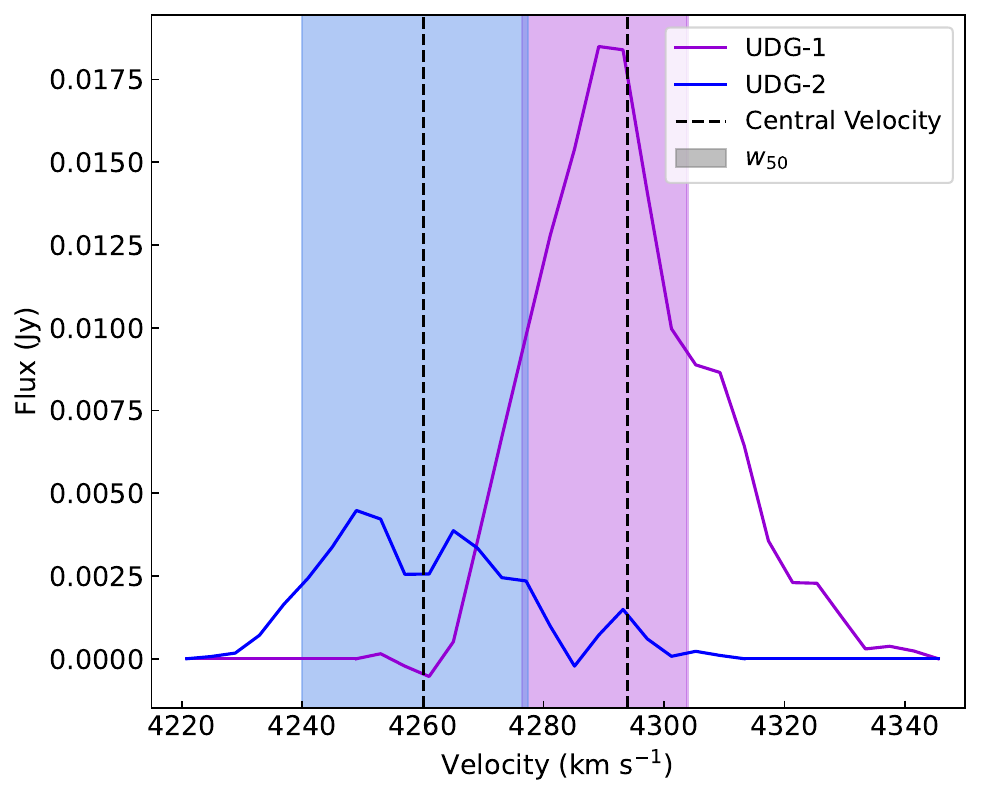}
         \caption{ }
         \label{fig:spec12}
     \end{subfigure}

     \begin{subfigure}[t]{0.47\textwidth}
         \centering
         \includegraphics[width=\textwidth]{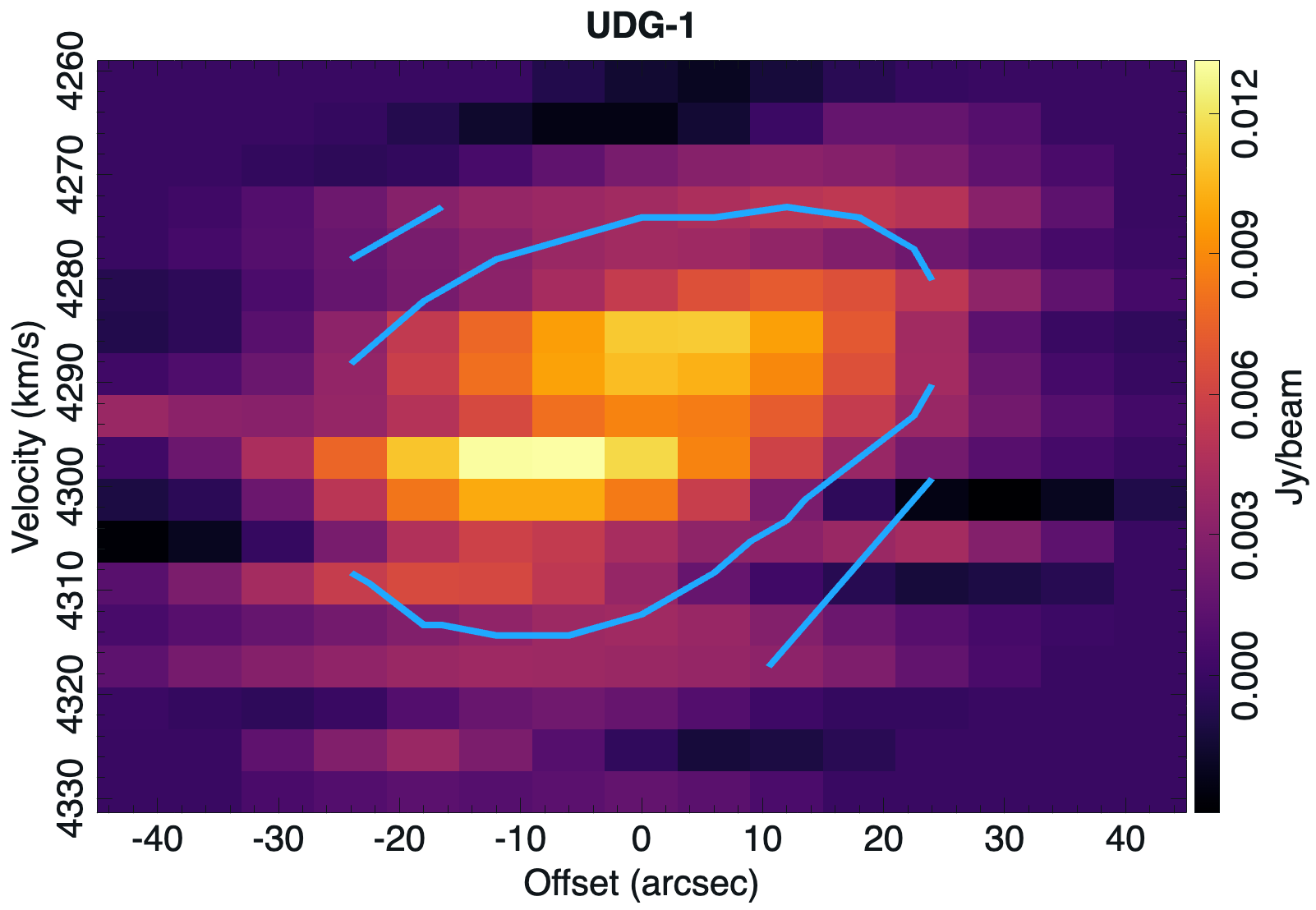}
         \caption{  }
         \label{fig:pv1}
     \end{subfigure}
     \hfill
     \begin{subfigure}[t]{0.48\textwidth}
         \centering
         \includegraphics[width=\textwidth]{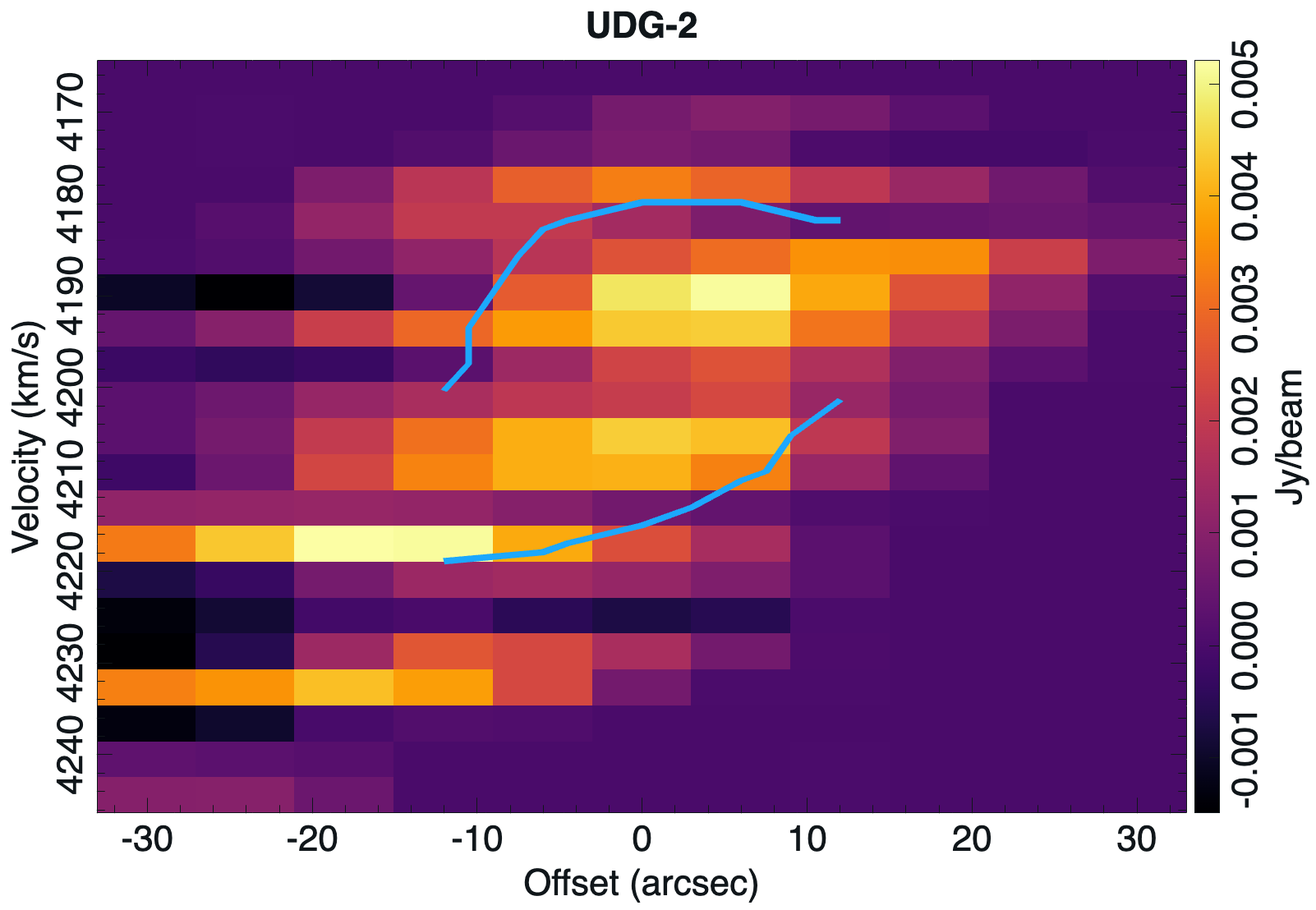}
         \caption{  }
         \label{fig:pv2}
     \end{subfigure}
    
    \caption{ H{\sc i} properties of the UDG pair. (a) The moment 0 (intensity) map. The H{\sc i} contours levels equal to the local column density rms ($2\times10^{19}$ cm$^{-2}$), $1\times10^{20}$ cm$^{-2}$ and $2\times10^{20}$ cm$^{-2}$ are overlaid. The dashed contour represents the edge of the SoFiA mask. Apertures that separate the H\,{\sc i} emission of each UDG are shown in cyan. (b) The moment 1 (velocity) map. The same H{\sc i} column density contours from (a) are shown. The axes used for the position velocity diagrams are shown in red. (c) The WALLABY masked spectrum (black) and unmasked spectrum (grey) measured in a 23$\times$23 pixel box. The central velocity are shown by vertical dashed lines and {$w_{20}$ }is shown by the shaded regions. (d) The spectra for UDG-1 (purple) and UDG-2 (blue) measured from the masked cube in the H\,{\sc i} apertures from (a). The central velocities are shown by vertical dashed lines and {$w_{20}$} are shown by the shaded regions. (e) Position-velocity diagram of UDG-1 along the axes shown in (b). {Blue} contours show [0.0019, 0.0037] Jy/beam ([1, 2]$\times$ the local rms). (f) Position-velocity diagram of UDG-2 along the axes shown in (b). {Blue} contours show 0.0019 Jy/beam (the local rms).  }
    \label{fig:hi}
    
\end{figure*}

We inspect the H\,{\sc i} properties and kinematics of the UDG pair. The integrated intensity map (moment 0 map) is shown in Figure \ref{fig:mom0}. The moment 0 map shows two peaks in the H{\sc i} emission which correspond to the locations of UDG-1 and UDG-2. To measure the mass of the H\,{\sc i} gas associated with each UDG, we use the apertures shown in the figure. As is the case with the stellar mass and SFR, the H\,{\sc i} mass of UDG-1 is larger than UDG-2. The H\,{\sc i} mass of UDG-1 is $\log_{10}(M_{HI}/\rm M_{\odot})=8.95$, and the H\,{\sc i} mass of UDG-2 is $\log_{10}(M_{HI}/\rm M_{\odot})=8.60$. 

Figures \ref{fig:spec} shows the SoFiA masked spectra and unmasked spectra of the entire system. The unmasked spectrum was measured in the Cube Analysis and Rendering Tool for Astronomy \citep[CARTA;][]{Comrie:21} in a 23x23 pixel box (major axis of the source is approximately 23 pixels) and has been baseline corrected (fit with a linear equation). The central velocity of the system is 4287 km s$^{-1}$  (1.400378 GHz). The $w_{50}$ is 22 km s$^{-1}$ (103 kHz) and the {$w_{20}$ is 64 km s$^{-1}$ (301 kHz).} Additionally, in Figure \ref{fig:spec12} we show the spectra for each UDG individually measured from the SoFiA masked cube in the H\,{\sc i} apertures from Figure~\ref{fig:mom0}. {We measure the central velocities of UDG-1 and UDG-2 to be 4294 km s$^{-1}$ and 4260 km s$^{-1}$, respectively. For UDG-1, the $w_{50}$ and $w_{20}$ are 27 km s$^{-1}$ and 48 km s$^{-1}$, while for UDG-2 they are 38 km s$^{-1}$ and 48 km s$^{-1}$, respectively }

Figure \ref{fig:mom1} presents the velocity field (moment 1 map) and Figures \ref{fig:pv1} and \ref{fig:pv2} show the position-velocity diagrams for each UDG along the axes shown in Figure \ref{fig:mom1}. {Contours corresponding to 1–2 times the local rms are shown in blue.} Both UDGs show evidence of a velocity gradient which could indicate rotation, however, higher resolution data is required to confirm this.

\subsection{Environment}
\label{sec:enviro}


\subsubsection{Location}

\begin{figure}
     \centering
     
       \begin{subfigure}[t]{0.47\textwidth}
         \centering
         \includegraphics[width=\textwidth]{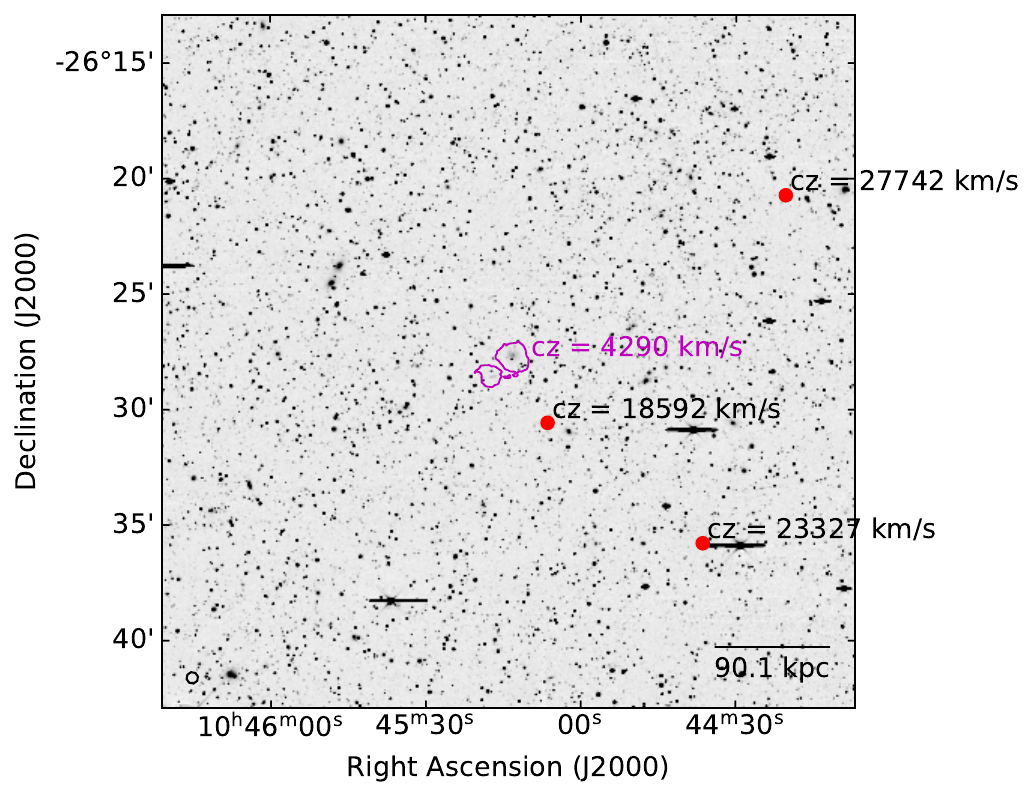}
         \caption{ }
         \label{fig:small}
     \end{subfigure}
     
     \begin{subfigure}[t]{0.47\textwidth}
        \centering
         \includegraphics[width=0.8\textwidth]{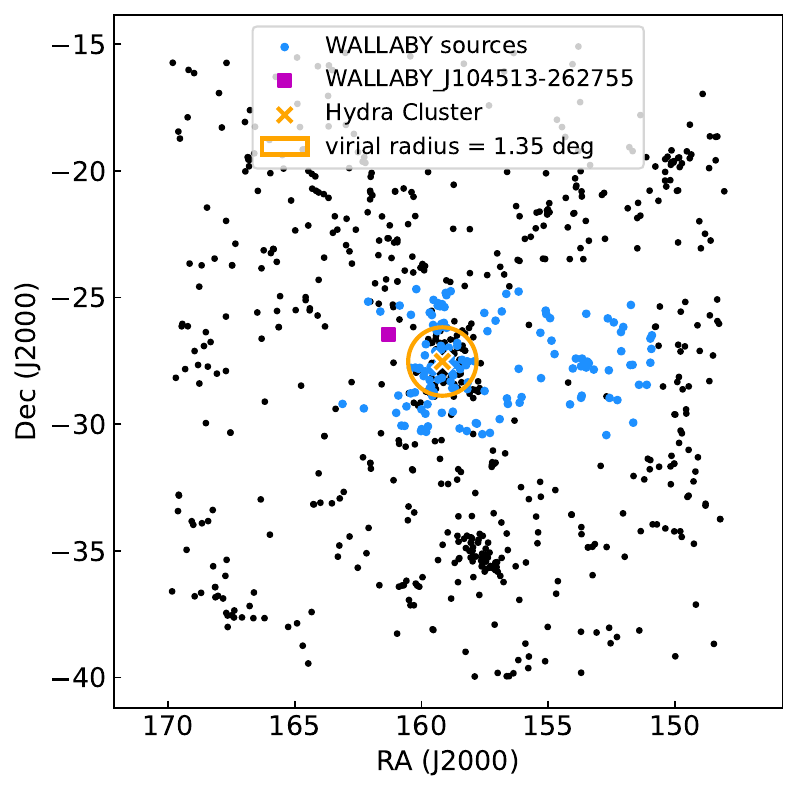}
         \caption{}
         \label{fig:large}
     \end{subfigure}

\caption{ {Location of the UDG pair. } (a) Legacy Survey g-band image centred on the UDG pair (H{\sc i} mask outline shown) with sources from NED that lie within a 15 arcmin radius highlighted. (b) The location of the UDG pair (purple square marker) with respect to the larger scale structure. The Hydra cluster and virial radius is shown in orange. The other WALLABY sources and sources from {6dFGS, 2MRS and MILLIQUAS} with redshifts $<0.015$ are shown in blue and black respectively. The UDG pair is 2.17 deg in projection from the cluster centre with a line of sight velocity difference of $\sim600$ km s$^{-1}$. }
    \label{fig:enviro}
    
\end{figure}

In this section we look at the small-scale local environment of WALLABY J104513-262755 as well as its location with respect to the larger-scale cosmological structure. Figure \ref{fig:small} shows the local environment around the UDG system. Overlaid onto the Legacy Survey g-band image are the locations and velocities of sources catalogued in the NASA/IPAC Extragalactic Database (NED) within 15 arcmin and have $z<0.1$. All of these nearby sources have velocities at least 14,000 km s$^{-1}$ larger than the central velocity of the UDG system, and consequently are background sources. In fact, there are no other {known} galaxies within a 30 arcmin radius (525 kpc at 61.9 Mpc) and $\pm 1000$ km s$^{-1}$, with the nearest galaxy, ESO 501-G082, having a separation of 30.3 arcmin and 240 km s$^{-1}$. {To further quantify the local isolation, we follow \cite{Verley:07} and estimate the local number density, $n_k$, and {the ratio between the tidal forces and binding forces,} $Q$. {$\log(n_k) \propto \log((k-1)/V)$} with $V=4\pi r_k^3/3$, where $k=5$ and $r_k$ is the projected distance to the 5$^{th}$ nearest neighbour in arcminutes. $Q_{ip} \propto (\sqrt{D_pDi}/S_{ip})^3$, where $D_p$ and $D_i$ are the diameters of the primary galaxy and i$^{th}$ neighbour respectively, $S_{ip}$ is the projected separation at the distance of the primary galaxy, and {$\log(Q) = \log(\Sigma_i Q_{ip})$.} Considering neighbouring galaxies within $\pm 1000$ km s$^{-1}$ and projected separation $< 1^{\circ}$, we find that {$\log(n_5)\sim-4.9$} and {$\log(Q) \sim -3.8$.} Based on Fig. 6 of \cite{Verley:07}, this places the UDG pair in an isolated environment, with similar (or even smaller) values to those of the isolated galaxies in the Analysis of the interstellar Medium of Isolated GAlaxies (AMIGA) sample, and with much smaller values than those of galaxy triplets, clusters, Hickson compact groups. } Hence we can conclude that WALLABY J104513-262755 is in a low density local environment. This suggests that its evolution is likely to have been driven by internal processes and interactions within the galaxy pair rather than the external environmental conditions thus far.

Figure \ref{fig:large} shows the location of WALLABY J104513-262755 in the context of the large-scale structure using data from a {compilation of literature redshifts from the 6dF Galaxy Survey \citep[6dFGS,][]{Jones:09}, 2MASS Redshift Survey \citep[2MRS,][]{Huchra:12}, the Million Quasars catalogue \citep[MILLIQUAS,][]{Flesch:21} and the WALLABY catalogue.} All galaxies with redshifts $<0.015$ are shown and the redshift of WALLABY J104513-262755 is 0.014.  We can see that WALLABY J104513-262755 lies outside the virial radius of the Hydra cluster, {at $\sim 1.6$ virial radii}, yet still relatively nearby (2.17 deg, or 2.2 Mpc at a distance of 61.9 Mpc, in projection from the cluster centre with a line of sight velocity difference of 601 km s$^{-1}$).  \cite{Reynolds:22} investigate properties of WALLABY galaxies falling into the Hydra cluster, however WALLABY J104513-262755 is excluded from their sample as they limit their selection to H{\sc i} sources with a single optical counterpart.  Placing WALLABY J104513-262755 onto the phase space diagram in Figure 2 of \cite{Reynolds:22}, we find that WALLABY J104513-262755 is in the region of infalling galaxies ($r/R200 = 1.4$, $\Delta v / \sigma_{disp}=0.97$). We note that \cite{Reynolds:22} adopt different Hydra cluster parameters from those we have chosen {(introduced at the beginning of Section \ref{sec:pair})} to create their phase space diagram. However, this does not change the conclusion that WALLABY J104513-262755 lies in the region of infalling galaxies.  We could be observing the UDG pair during a crucial stage of its evolution, seeing its cold gas reservoir before it experiences significant ram pressure stripping as it enters the Hydra cluster.

{
\cite{delacasa:25} study the gas and star forming properties of galaxies within and surounding the Hydra cluster out to ~1.75 $R_{200}$. Although the UDG pair is too low surface brightness to be included in their sample, it appear to be infalling from the same direction as subgroup 4, which is thought to be accreting from the filament that connects the Hydra cluster to the Shapley supercluster. \cite{delacasa:25} find that the most recent infall into the Hydra cluster consists of galaxies associated with this filament as well as with the filament shared with Antlia. Ram pressure stripping has been shown to affect galaxies out to 2-3 $R_{200}$ \citep{zinger:18}, and the H{\sc i} contours of UDG-2 could suggest this pair is also influenced. However, given the low signal-to-noise of the relevant contours and the very small separation of the pair, it is not possible to disentangle any ram pressure effects from tidal interactions, which are likely dominate.}


\subsubsection{Interactions}
\label{sec:int}

{Next} we examine the potential interactions taking place between UDG-1 and UDG-2, as well as those between the UDG pair and the Hydra cluster. The projected separation between the optical centres of the pair is just 75 arcsec (equivalent to 22 kpc at a distance of 61.9 Mpc), with a central H\,{\sc i} velocity difference of 34 km s$^{-1}$. \cite{Pfister:20} use the \textsc{HORIZON AGN} simulation to investigate selection criteria for galaxy pairs that will undergo mergers. They find that their best thresholds for merging galaxies are those with projected separations $<100$ kpc and redshift difference $<0.001$ {($\Delta cz < 300$~km~s$^{-1}$)}. These criteria indicate that the pair may merge in the future. 

 {
Investigating ongoing interactions in more detail requires estimating the dynamical masses. The dynamical mass ($M_{dyn}$) enclosed in radius $R$ (assuming a rotation-supported, spherically symmetric system) is given by $M_{dyn} = v_{circ}^2 R / G$, where $v_{circ}$ is the {rotational} velocity and $G$ is the gravitational constant. Unfortunately, as the H{\sc i} gas is only marginally resolved, we are unable to model the rotation curves of each UDG to measure their {rotational} velocities. Instead, we can estimate the {rotational} velocities from their measured {$w_{20}$} emission line widths, correcting for the effect of inclination {($v_{circ}=w_{20}/(2\sin(i))$)}. We estimate that UDG-1 and UDG-2 have {rotational} velocities of { 39.9 km~s$^{-1}$ and 24.2 km~s$^{-1}$, which gives rise to dynamical masses (within the radii of the H{\sc i} apertures) of $\sim6\times10^9$~M$_{\odot}$ and $\sim2\times10^9$~M$_{\odot}$ respectively. } Alternatively, the {rotational} velocities could also be estimated using the baryonic Tully-Fisher (bTF) relation from \cite{McGaugh:00}. Using this method, we estimate that the {rotational} velocities of UDG-1 and UDG-2 are 79 km s$^{-1}$ and 64 km s$^{-1}$ respectively. Subsequently, their dynamical masses are significantly larger at $\sim2\times10^{10}$ M$_{\odot}$ and $\sim1\times10^{10}$ M$_{\odot}$ respectively. Due to the diverse nature of the UDG population, it remains unclear whether most gas-rich UDGs are expected to follow the bTF relation or deviate from it \citep[e.g.][]{mancerapina:19, Lelli:24}. Using the {rotational} velocities derived from the inclination-corrected {$w_{20}$}, and assuming the optical inclinations are accurate, UDG-1 and UDG-2 would need to be located at a distance that is approximately {5} times closer in order to lie on the baryonic Tully-Fisher relation.}  The dynamical masses estimated from {rotational} velocities obtained through these two methods are inconsistent, highlighting the importance of high resolution H{\sc i} data and accurate kinematic modelling. Consequently, the strength of the interactions taking place will depend on the {rotational} velocities adopted, which we will explore throughout the rest of this subsection.

 {
To quantify and compare the strength of the tidal interactions between UDG-1 and UDG-2, as well as the UDG pair and the Hydra cluster, we can use the tidal strength parameter {\citep{Oh:08,wang:22}} outlined in Equation \ref{eq:stid} :
\begin{equation}
    S_{tid} = \left( \frac{M_{p}}{M_{g}} \right) \left( \frac{R_g}{d_{proj}} \right)^2 \left( \frac{v_{circ}}{\sqrt{(\Delta v_{rad})^2 + v_{smooth}^2}} \right)
    \label{eq:stid}
\end{equation}
where $M_p$ and $M_g$ are the dynamical masses of the perturber and galaxy respectively, $R_g$ is the galaxy radius, $d_{proj}$ is the projected distance between the galaxy centres, $v_{circ}$ is the {rotational} velocity of the galaxy, $\Delta v_{rad}$ is the difference in the line of sight velocities between the two objects and $v_{smooth} = 0$ is the smoothing velocity which can be included to avoid divide by zero errors.
}

 {
Table \ref{tab:Stid} presents the $S_{tid}$ values derived using the two methods for estimating the {rotational} velocities. From left to right, the columns show the tidal strength of UDG-1 on UDG-2 (setting UDG-1 as the perturber and UDG-2 as the effected galaxy), the tidal strength of UDG-2 on UDG-1, and the tidal strength of the Hydra cluster on the UDG pair. Regardless of the {rotational} velocities adopted, the tidal force of UDG-1 on UDG-2 is greater that that of UDG-2 on UDG-1, and consequently we expect {that} UDG-2 could be affected by the presence of UDG-1. This is unsurprising as the stellar and H\,{\sc i} masses of UDG-1 are larger than those of UDG-2. However, the strength of the tidal interactions ongoing within the pair compared to the effect of the Hydra cluster is very dependent on the dynamical masses. When adopting the dynamical masses derived from the {$w_{20}$} emission line width, UDG-1 experiences a similar tidal force from the Hydra cluster to its companion. On the other hand, when adopting the dynamical masses derived from the bTF relation, the UDGs in the pair appear to be much more tightly bound to each other.
}
\begin{table}
    \centering
    \caption{The tidal strength values ($S_{tid}$, Equation \ref{eq:stid}) derived using the two methods for estimating the {rotational} velocities (inclination corrected {$w_{20}$} and the baryonic Tully-Fisher relation). From left to right, the columns show the method, the tidal strength of UDG-1 on UDG-2 (setting UDG-1 as the perturber and UDG-2 as the {affected} galaxy), the tidal strength of UDG-2 on UDG-1, and the tidal strength of the Hydra cluster on the UDG pair. }
    \label{tab:Stid}
    \setlength{\tabcolsep}{3.5pt}
    \begin{tabular}{cccc}
	\hline
       Method  &  Perturber: UDG-1 & Perturber: UDG-2 & Perturber: Hydra cluster \\
	\hline
       {$w_{20}$}  &  {0.77} & {0.17} & {0.38} \\
       bTF & 1.15 & 0.60 & 0.08 \\
	\hline
    \end{tabular}
    
\end{table}

 {
To further explore this effect, we investigate whether UDG-1 and UDG-2 are gravitationally bound to each other and the Hydra cluster following the methods of \cite{Wolfinger:16} and \cite{Brough:06}. A two-body system is bound if the kinetic energy is less or equal to the potential energy of the system. Hence, a 2 body system is bound if the following equation is satisfied:
\begin{equation}
    (\Delta v_{rad})^2d_{proj} \leq 2GM_{tot} ~\sin^2\alpha~\cos\alpha
    \label{eq:bound}
\end{equation}
where $\Delta v_{rad}$ is the difference in line of sight velocities, $d_{proj}$ is the projected distance, $G$ is the gravitational constant, $M_{tot} $ is the total mass of the system, and $\alpha$ is the projection angle that can vary between $0^{\circ}$ and $90^{\circ}$. {$\alpha$ is the angle between the plane of the sky and the line joining the centres of the two sources. } The factor of $\sin \alpha $ accounts for the projection effects in $\Delta v_{rad}$, and the factor of $\cos \alpha$ corrects for projection effects in $d_{proj}$.}  The probability {that} the system is bound can be found by integrating $\int \cos\alpha~d\alpha$ over the values of $\alpha$ that satisfy Equation \ref{eq:bound}. Table \ref{tab:bound} shows the range of projection angles for which the UDG-pair are bound to each other and the Hydra cluster, considering the two methods for estimating {rotational} velocities. Figures to help visualise this effect can be found in Appendix \ref{appen:int}. The UDG pair remain bound to the Hydra cluster for $41^{\circ}<\alpha<68^{\circ}$ regardless of the dynamical masses used, as the mass of the Hydra cluster is significantly larger. {This range corresponds to a probability of 27 per cent.} However, again we find that this does affect how we view the interactions ongoing within the pair. When using the larger dynamical masses derived from the bTF relation, the pair is bound for $18^{\circ}<\alpha<85^{\circ}$ ({corresponding to a probability of 68 per cent)}. In contrast if we adopt the smaller dynamical masses derived from the inclination corrected {$w_{20}$} emission line widths, the pair appears to be completely unbound {for all $\alpha$, though only marginally.}

\begin{table}
    \centering
    \caption{The range of projection angles for which the UDG-pair are bound to each other and the Hydra cluster following \protect\cite{Wolfinger:16}, considering the two different methods for estimating {rotational} velocities (inclination corrected {$w_{20}$} and the baryonic Tully-Fisher relation). }
    \label{tab:bound}
    \begin{tabular}{ccc}
	\hline
       Method  &  UDG-1 and UDG-2 & UDG pair and Hydra cluster \\
	\hline
       {$w_{20}$}  & unbound & $41^{\circ}<\alpha<68^{\circ}$ \\
       bTF & $18^{\circ}<\alpha<85^{\circ}$ & $41^{\circ}<\alpha<68^{\circ}$ \\
	\hline
    \end{tabular}
    
\end{table}

We’ve demonstrated that the choice of {rotational} velocity (and the subsequent dynamical mass) plays a crucial role in our interpretation of the interactions taking place. It highlights the utility of high-resolution H{\sc i} data, and the caution one must exercise when drawing conclusions from marginally resolved data. Despite these challenges, this analysis suggests that the system may experience significant dynamical changes in the future. UDG-2 could be perturbed by UDG-1, with the possibility of an eventual merger.  {The presence of the Hydra cluster will likely increase the merger timescale relative to an similar pair evolving in complete isolation.} The pair may be infalling into the Hydra cluster, where it could lose its cold gas reservoir through ram pressure stripping and join the other gas-poor UDGs already within the cluster. 


There are notable parallels between the UDG-1/UDG-2 system, and the Large and Small Magellanic Clouds (LMC/SMC). Like our UDG pair, both the LMC and SMC possess similar H{\sc i} masses \citep[][]{Stanimirovic:04,Staveley-Smith:03}, and just as UDG-1 has a larger stellar mass than its companion, so does the LMC \citep{vanderMarel:02,Rubele:18}. The LMC and SMC have a very small separation \citep[$\sim20$ kpc and 103 km s$^{-1}$;][]{Zivick:18}, comparable with that of the UDG pair. {Galaxies with shallow potential wells can develop off-centre bars, and} the LMC exhibits a bar structure that may have been tidally induced \citep{Besla:12}, a scenario that could similarly apply to UDG-1. Although the UDG pair currently shows no signs of tidal debris comparable to the Magellanic bridge, it would be interesting to see whether deeper observations could reveal the presence of diffuse tidally stripped gas. Additionally, follow-up high resolution H{\sc i} observations are necessary to fully understand the kinematics of this dynamic system.


\section{Comparison with Other UDGs}
\label{sec:comparison}

In this section we compare the UDG pair to the other WALLABY UDGs, the (gas-poor) UDGs in the Hydra cluster as well as other H{\sc i}-bearing UDGs {observed in low density environments and in simulations}.

\subsection{A UDG-LSBG pair in WALLABY}
\label{sec:otherpair}

\begin{figure}
    \centering
    \includegraphics[width=0.97\linewidth]{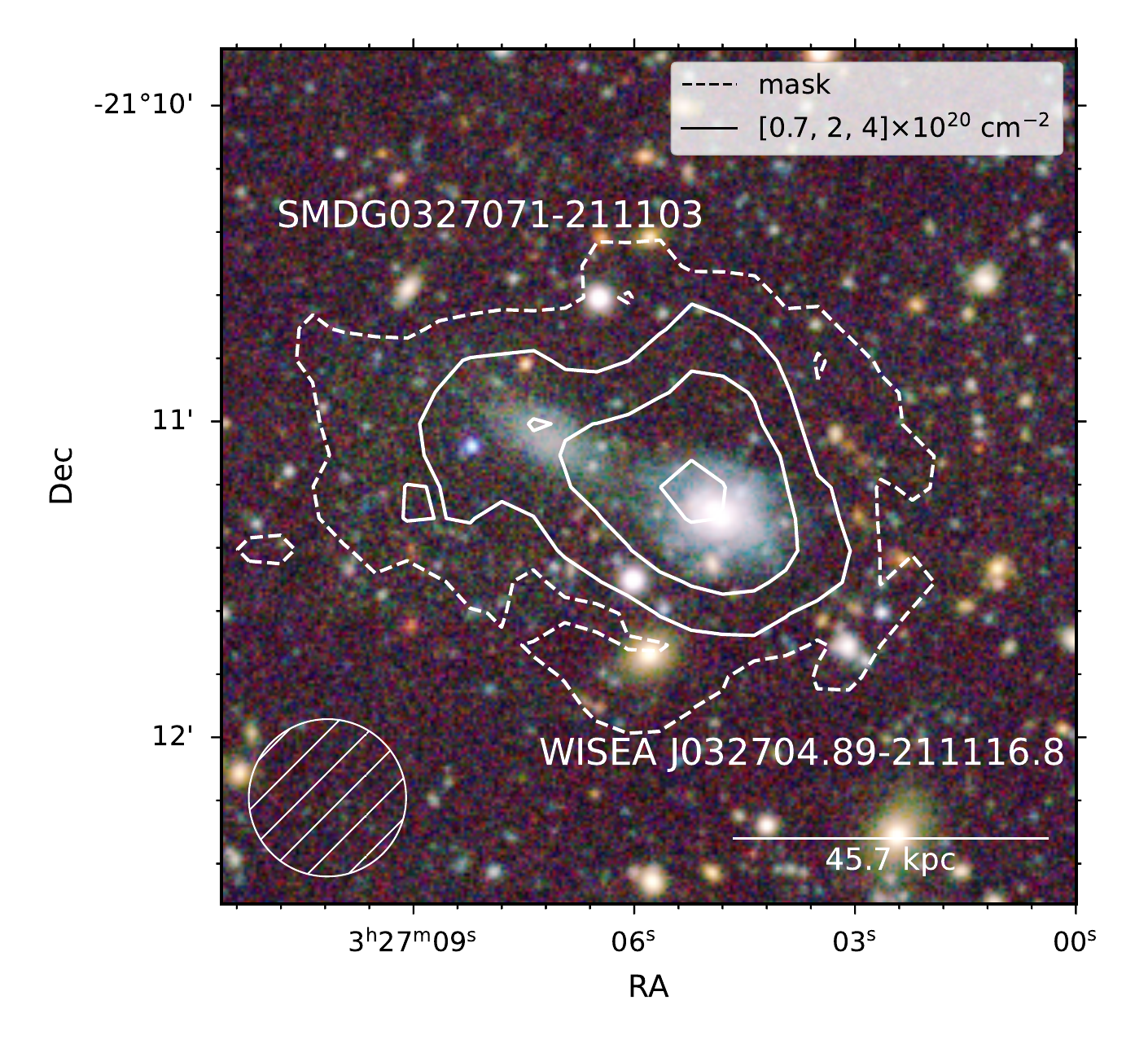}
    \caption{ The three colour ($grz$-band) Legacy Survey image of WALLABY J032705.80-211110.4. The H{\sc i} contour level equal to the local column density rms ($7\times10^{19}$), $2\times10^{20}$ and $4\times10^{20}$ cm$^{-2}$ are overlaid in white. The dashed contour represents the edge of the SoFiA mask. The WALLABY detection contains two galaxies visible in the Legacy Survey image: the UDG, SMDG0327071–211103, and the LSBG, WISEA J032704.89–211116.8. The WALLABY beam (30 arcsec) is shown in the lower left corner.}
    \label{fig:otherpair}
\end{figure}

{Our UDG sample includes not only the UDG-UDG pair WALLABY J104513-262755, discussed in detail in the previous section, but also the UDG-LSBG pair WALLABY J032705.80-211110.4, shown in Figure \ref{fig:otherpair}.  The WALLABY detection contains two galaxies visible in the Legacy Survey image: SMDG0327071–211103, a UDG with $\mu_{0,g} = 25.0$ mag arcsec$^{-2}$ and $R_e = 5.9$ kpc; and its companion WISEA J032704.89–211116.8, with $\mu_{0,g} = 20.8$ mag arcsec$^{-2}$ and $R_e = 4.2$ kpc. The latter has a mean $g$-band surface brightness within 1 $R_e$ of 23.3 mag arcsec$^{-2}$, meeting the LSBG criteria used by \cite{obeirne:25}, but too bright to be considered a UDG. Based on the central coordinates of the Sérsic models, the UDG-LSBG pair has a projected separation of just 35 arcsec, corresponding to 25 kpc at 157 Mpc,  comparable to the separation between UDG-1 and UDG-2 (22 kpc). This pair is further away that UDG-1 and UDG-2, and is slightly worse resolved, ($\sim 4$ beams across the major axis), and unlike the UDG-UDG pair, this UDG-LSBG pair has a common H{\sc i} envelope. The peak H{\sc i} emission occurs at the LSBG companion, and the total H{\sc i} mass and $w_{50}$ of the UDG-LSBG pair is $\log M_{\rm HI}/M_\odot = 9.7$ and 151 km s$^{-1}$, compared with $\log M_{\rm HI}/M_\odot = 9.0$ and 22 km s$^{-1}$ for the UDG–UDG pair.  Just like the UDG-UDG pair, this pair is also relatively isolated. The only galaxy within a 1 degree radius and $\pm 1000$ km s$^{-1}$ is WISEA J032910.29-220116.3,  with a separation of 58 arcmin and a velocity difference of 244 km s$^{-1}$. With the complete WALLABY survey, we will be able to probe the frequency of H{\sc i}-bearing UDG–UDG and UDG–LSBG pairs, which will allow us to learn more about the formation and evolution of these extreme galaxies.}

\subsection{UDGs in the literature}
{Simulations have shown that H{\sc i}-bearing UDGs with properties consistent with observations preferentially form in low-density environments \citep[e.g.][]{Wright:21,Benavides:23}. They also provide insight into the fate of gas-rich UDGs near clusters and overdensities. Using zoom-in cosmological simulations, \citep{Jiang:19} found that $\sim$20 per cent of field UDGs that fall into a massive halo continue to survive as satellite UDGs and experience ram pressure stripping. Among UDGs found in and around massive haloes, those located toward the inner regions are almost always quiescent and red, whereas those beyond the virial radius exhibit a wider range of colours.}

There are 22 UDGs that meet the \cite{vanDokkum:15} criteria (within errors) that have been detected within $0.4R_{\rm vir}$ of the Hydra cluster \citep{Iodice:20,LaMarca:22,Iodice:23}. They were initially observed in the VST Early-type Galaxy Survey (VEGAS) and are being followed up as part of the Looking into the faintEst WIth MUSE (LEWIS) program. Using the UDG abundance-halo mass relation from \cite{vanderBurg:17}, \cite{Iodice:23} found the expected number of UDGs within the virial radius of the Hydra cluster to be $48\pm10$. The cluster UDGs have been shown to have similar structural and photometric properties and globular cluster content to other low surface brightness galaxies and dwarfs in the cluster. This suggests they may represent the extreme low surface brightness tail of the size-luminousity distribution of dwarf galaxies in the cluster environment \citep{LaMarca:22}.  Using the phase-space diagram, \cite{Forbes:23} found that the UDGs are similarly concentrated to the other cluster galaxies. They also find that many of the UDGs had early infall times ($>6.45$ Gyr) and likely had experienced early quenching.

The ALFALFA survey has been shown to be a powerful tool for identifying HUDs. Across the $\sim 7000$ deg$^{2}$ covered by the completed ALFALFA survey, \cite{Janowiecki:19} identified 71 H{\sc i}-bearing UDGs that satisfy the \cite{vanDokkum:15} UDG criteria using optical imaging from the the Sloan Digital Sky Survey (SDSS) \citep{Abazajian:09}. This corresponds to $\sim 1$ UDG per 100 deg$^2$.  We have identified {10 UDGs within the $\sim 690$ deg$^2$ spanned by the WALLABY pilot and currently available full survey fields that overlap with the Legacy Survey footprint. This is equivalent to $\sim 1.4$ UDGs per 100 deg$^{2}$.} The slightly higher detection rate of WALLABY is likely due to its improved resolution, which reduces source confusion, and the greater depth of the Legacy Survey compared to SDSS. 

\cite{Kado-Fong:22} study the star formation properties in 22 HUDs selected from the \cite{Janowiecki:19} sample, { showing that HUDs have low star formation efficiencies for their atomic gas masses. At fixed H{\sc i} mass, their stellar mass-weighted sizes are comparable to those of other H{\sc i}-bearing dwarfs.} \cite{He:19} identify 11 additional edge-on HUDs in the ALFALFA $40\%$ survey after correcting for the viewing angle. {These constitute 13 per cent of the total ALFALFA UDG sample and exhibit properties consistent with the face-on HUDs.} SMUDGes is a catalogue of 7070 UDG candidates, however without distance measurements they cannot be confirmed. \cite{Karunakaran:24} followed-up 378 of the UDG candidates, detecting H\,{\sc i} in 110 sources and confirming 37 of them as UDGs. {They find that the H{\sc i}-detected UDGs tend to have bluer colours, more irregular morphologies and more isolated environments than the H{\sc i} non-detected UDGs.}  Internal mechanisms are likely to be responsible for the extreme sizes and surface brightnesses, and environmental effects may be required to convert HUDs into gas-poor cluster UDGs \citep{Janowiecki:19}.

\begin{figure}
    \centering
    \includegraphics[width=\linewidth]{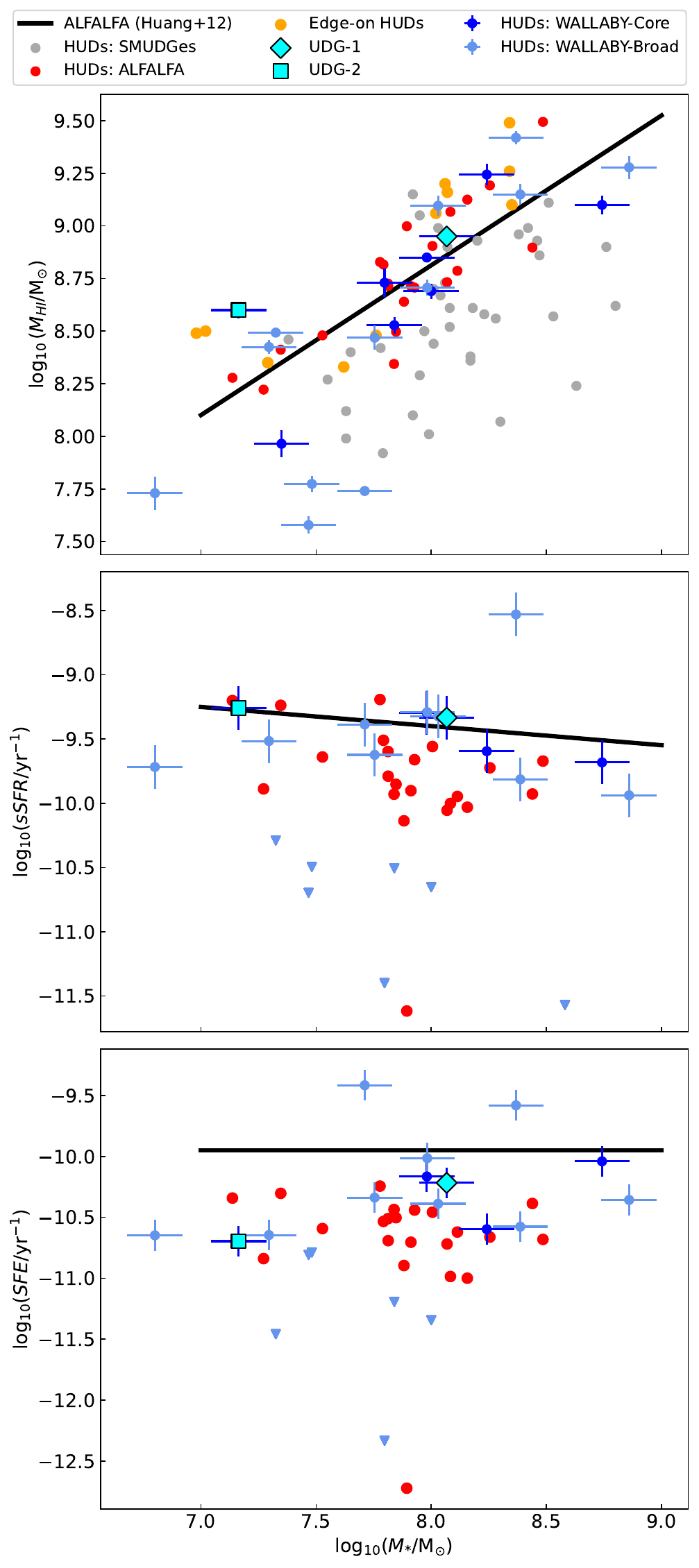}
    \caption{Comparing the properties of the UDG-pair and the WALLABY UDG sample with other galaxy populations. The upper, middle, and lower panels show H I mass, sSFR, and SFE as functions of stellar mass, respectively. The ALFALFA scaling relations are shown by the black lines. SMUDGes HUDs are shown in grey, ALFALFA UDGs from Kado-Fong et al. (2022) are shown in red, and edge-on ALFALFA UDGs from He et al. (2019) are shown in orange. The WALLABY UDGs are shown in blue, with the darker blue representing the {core} sample and the lighter blue representing the broader sample. Upper limits are shown by triangle markers. The blue diamond and square are used to represent UDG-1 and UDG-2 respectively.}
    \label{fig:scalerel}
\end{figure}

Figure \ref{fig:scalerel} compares the gas and star formation properties for our WALLABY UDGs to those from previous studies.  We show various ALFALFA relations \citep{Huang:12} with solid black lines, as ALFALFA is a H\,{\sc i}-selected sample with a similar sensitivity to WALLABY. The top panel shows the H{\sc i} mass as a function of the stellar mass. The ALFALFA HUDs, the UDG-pair and most of the WALLABY UDGs closely follow the ALFALFA stellar mass-H\,{\sc i} mass relation. This suggests that the UDG-pair and the majority of WALLABY UDGs are consistent with the broader H\,{\sc i}-selected galaxy population in the low stellar mass regime. On the other hand, the SMUDGes galaxies appear H\,{\sc i} deficient for their stellar mass, which is likely the result of selection effects. 

The middle panel shows the specific SFR (sSFR; star formation rate divided by stellar mass) as a function of the stellar mass. Some of the ALFALFA UDGs follow the ALFALFA stellar mass-sSFR relation, while the rest have low sSFRs. This same trend appears with the rest of the WALLABY UDGs, except one from the broader UDG sample with a notably larger sSFR. Both UDG-1 and UDG-2 follow the ALFALFA scaling relation closely {and have relatively high sSFR compared to the UDG population. It is unclear whether galaxy interactions or the high H{\sc i} mass fraction are responsible for these high sSFRs}.  The bottom panel presents the star formation efficiency (SFE; star formation rate divided by H{\sc i} mass) as a function of the stellar mass, with the mean SFE of the ALFALFA galaxies shown by the black line. Some of the WALLABY UDGs, including UDG-1, have a SFE very similar to that of the ALFALFA population, while most other WALLABY UDGs have SFEs at least $0.5$ dex lower.  There appears to be two types of WALLABY UDGs: those with the SFE of typical H{\sc i}-selected galaxies, and those that are inefficient at converting the H{\sc i} gas to stars.

As UDG-1 and UDG-2 are consistent with the other HUDs, it suggests that  {any potential ongoing interactions} have not affected their properties significantly. {If the current SFRs in many of the gas-rich UDGs presented here persist over the next few gigayears, their stellar masses will increase by a factor of a few as their gas reservoirs are gradually depleted. As a result, their surface brightnesses may increase sufficiently for them to cross the UDG definition and enter into the `typical' galaxy population. This implies that gas-rich UDGs may not remain in the UDG phase indefinitely.}

\begin{figure*}
     \centering
       \begin{subfigure}[t]{0.3\textwidth}
         \centering
         \includegraphics[width=\textwidth]{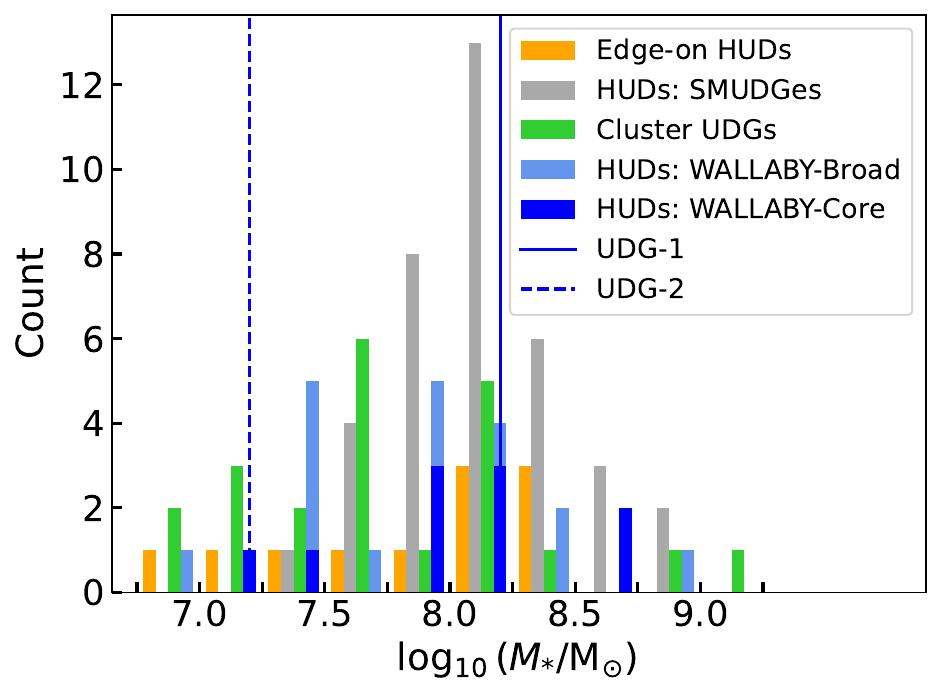}
         \caption{  }
         \label{fig:hmstar}
     \end{subfigure}
     \begin{subfigure}[t]{0.3\textwidth}
         \centering
         \includegraphics[width=\textwidth]{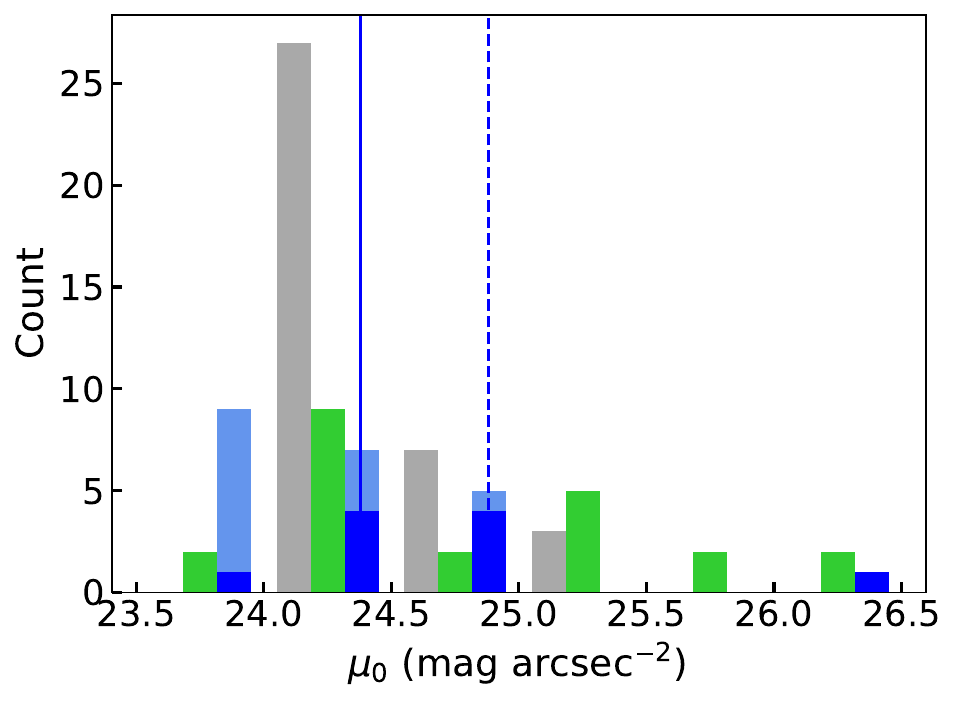}
         \caption{  }
         \label{fig:hmu0}
     \end{subfigure}
     \begin{subfigure}[t]{0.3\textwidth}
         \centering
         \includegraphics[width=\textwidth]{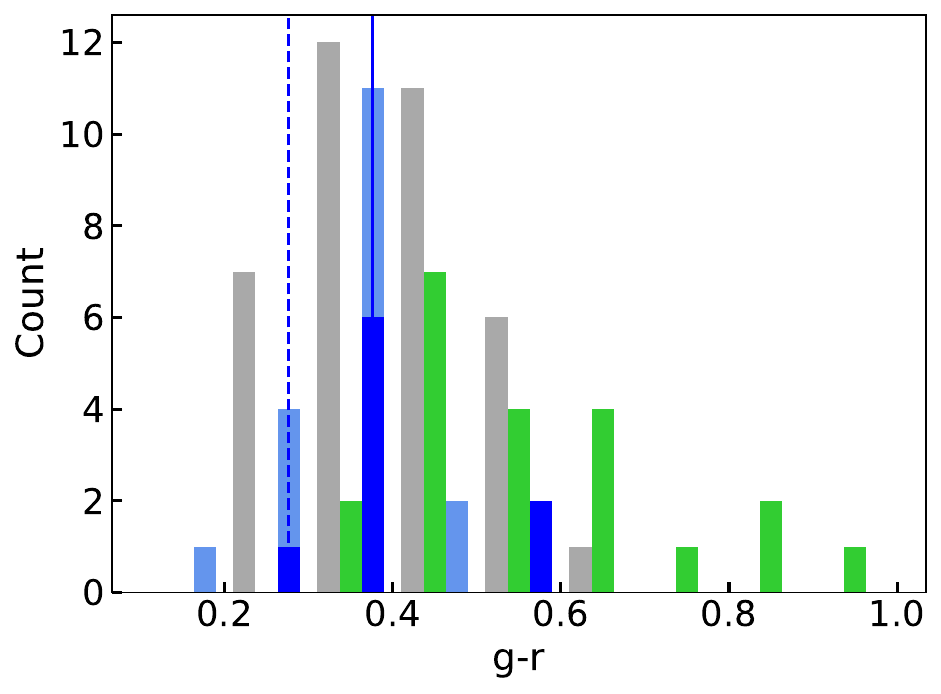}
         \caption{  }
         \label{fig:hgr}
     \end{subfigure}
           \begin{subfigure}[t]{0.3\textwidth}
         \centering
         \includegraphics[width=\textwidth]{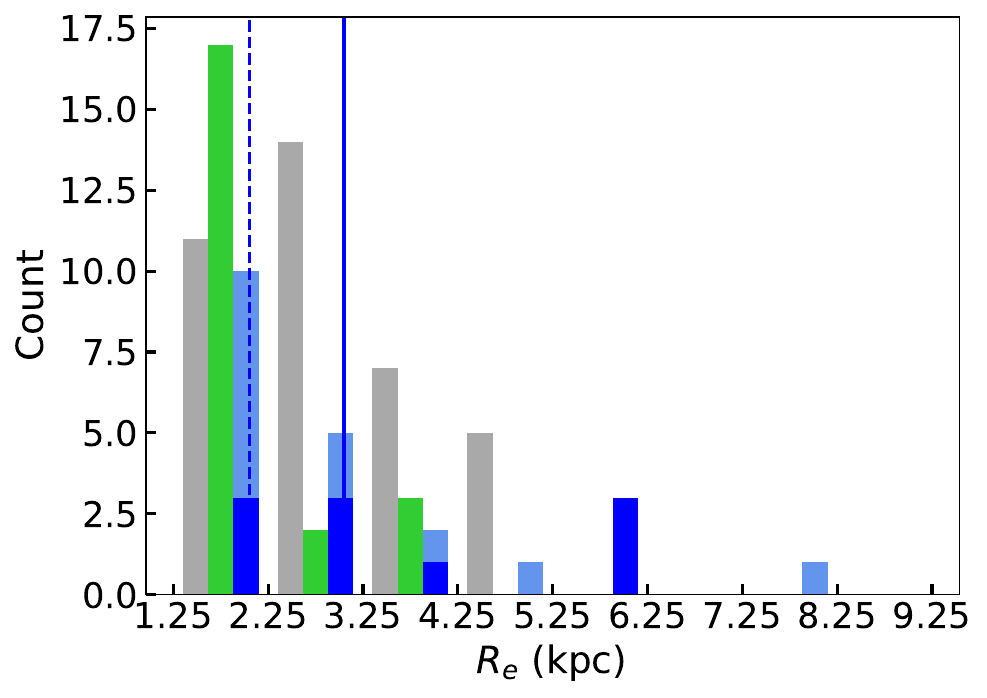}
         \caption{  }
         \label{fig:hRe}
     \end{subfigure}
     \begin{subfigure}[t]{0.3\textwidth}
         \centering
         \includegraphics[width=\textwidth]{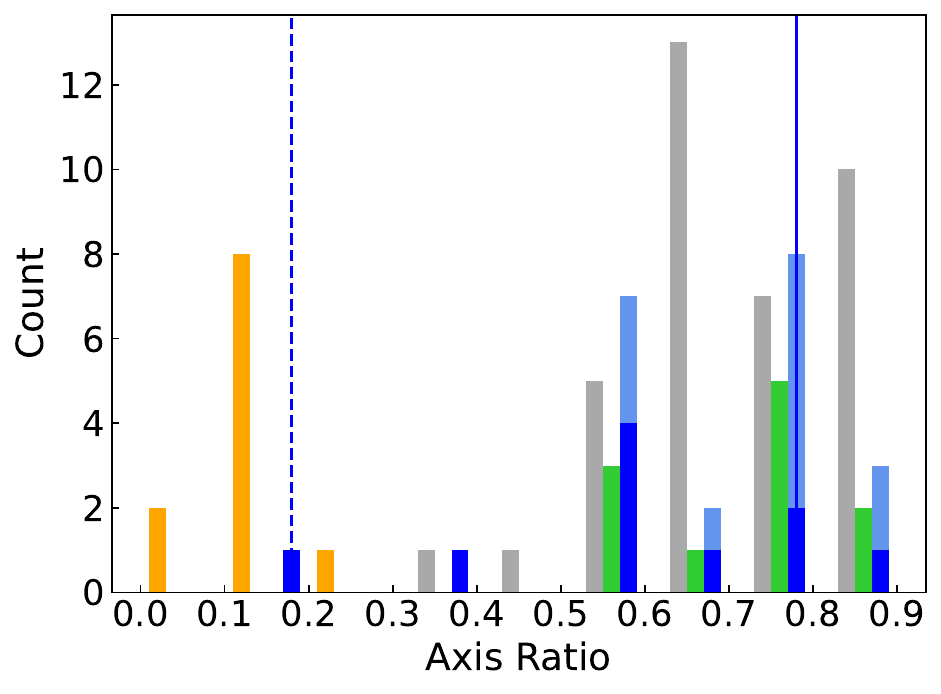}
         \caption{  }
         \label{fig:hq}
     \end{subfigure}
     \begin{subfigure}[t]{0.3\textwidth}
         \centering
         \includegraphics[width=\textwidth]{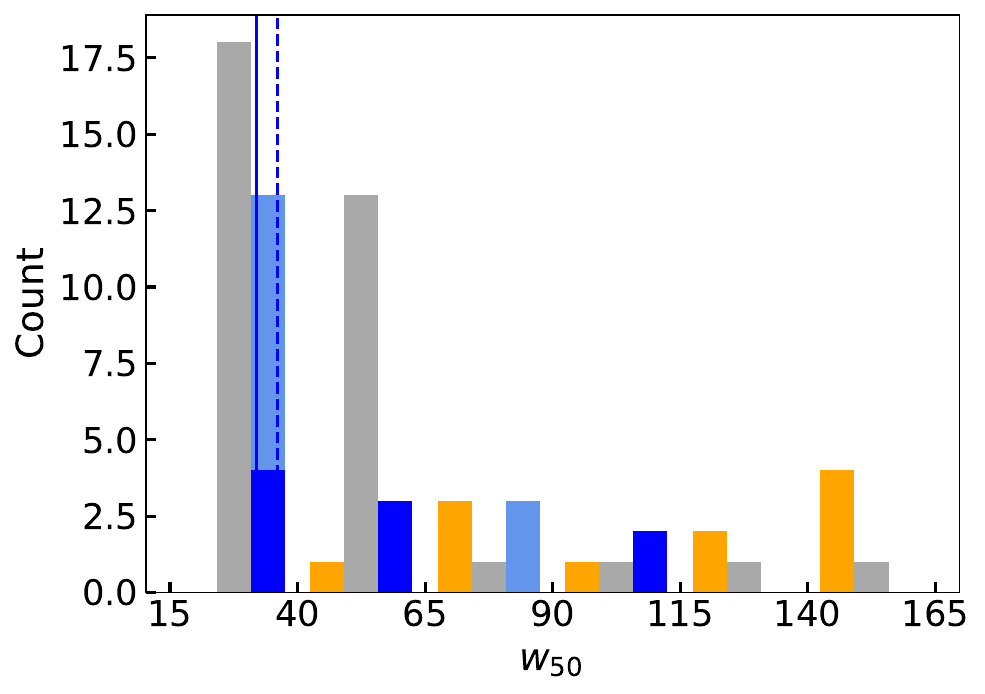}
         \caption{  }
         \label{fig:hw50}
     \end{subfigure}
     
    \caption{Properties of the WALLABY UDGs compared with other HUDs and UDGs in the Hydra cluster.  All 22 Hydra cluster UDGs are shown in green from \protect\cite{Iodice:20,LaMarca:22,Iodice:23}, however, axis ratios were only available for half the sample. The SMUDGes HUDs \protect\citep{Karunakaran:24} are shown in grey and the ALFALFA edge-on HUDs \protect\citep{He:19} are shown in orange. The WALLABY UDGs are shown in blue, and the vertical lines indicate which bins UDG-1 and UDG-2 are in. Histograms show: (a) stellar mass, (b) central $g$-band surface brightness, (c) $g$-$r$ colour, (d) effective radius, (e) axis ratio, and (f) {$w_{50}$} H\,{\sc i} emission line width.}
    \label{fig:hist}
\end{figure*}

Figure \ref{fig:hist} presents histograms of the properties of WALLABY UDGs for comparison with the other HUDs and the Hydra cluster UDGs. All 22 Hydra cluster UDGs are shown in green from \cite{Iodice:20}, \cite{LaMarca:22} and \cite{Iodice:23}, however, in Figure \ref{fig:hq}, axis ratios were only available for half the sample. The SMUDGes HUDs \citep{Karunakaran:24} are shown in grey and the ALFALFA edge-on HUDs \citep{He:19} are shown in orange. The WALLABY UDGs are shown in blue, and the vertical lines indicate which bins UDG-1 and UDG-2 are in. The WALLABY UDGs have bluer colours than the cluster UDGs, and are consistent with the HUDs sample.  The axis ratio of UDG-2 is significantly smaller than all of {the} axis ratios of the cluster UDGs, SMUDGes HUDs, and other WALLABY UDGs, suggesting that edge-on UDGs are unlikely to be detected outside of a dedicated search applying a correction for inclination, as was done by \cite{He:19}. {The axis ratio distribution of the H{\sc i}-bearing UDGs is inconsistent with that of the general dwarf and LSBG population. \cite{Staveley-Smith:92} show that, in those populations, the relative frequency of galaxies is expected to increase towards higher axis ratios following a power law trend.} The {$w_{50}$} emission line widths of the WALLABY galaxies, including UDG-2, are small like most of the SMUDGes galaxies, contrasting the edge-on HUDs.

Figure \ref{fig:hRe} highlights that there are {4} WALLABY UDGs with very large effective radii ($R_e > 5$ kpc) compared to UDGs in the literature. These sources are WALLABY J103508-283427, WALLABY J131403-120644, WALLABY J131845-153954 and {WALLABY J032705-211110 (the UDG from the UDG-LSBG pair)}. WALLABY J103508-283427 is a member of the interacting group Klemola 13 (see \citealt{obeirne:24} for further details). It has an effective radius of 6.1 kpc at a distance of 47.5 Mpc. The tidal interactions in its evolutionary history may be responsible for its extended effective radius, similar to what \cite{Duc:14} observed in tidal dwarf galaxies. WALLABY J131403-120644 and WALLABY J131845-153954 are two of the most distant UDGs in our sample (at 95.8 Mpc and 94.1 Mpc respectively) and have effective radii of 5.4 kpc and 7.9 kpc respectively. These sources had not been previously studied, and so no distance measurements, other than the luminosity distance we infer from the H{\sc i} velocities, are available. If these sources were closer, for example at the distance of the UDG-pair (61.9 Mpc), their effective radii would measure 3.5 kpc and 5.3 kpc. Both of these sources are marginally resolved (approximately 3 beams across their major axes) and there is some evidence of rotation in the moment 1 map of WALLABY J131403-120644 (see Figure \ref{fig:a3}). The large effective radii of these sources make them extreme, even for UDGs. They share similarities with `Nube', an almost dark galaxy with $\langle \mu_V\rangle_e \sim 26.7$ mag arcsec$^{-2}$ and a stellar mass of $\sim 4\times10^8$ M$_{\odot}$ \citep{Montes:24}. Nube is potentially associated with an H{\sc i} counterpart, which would suggest the galaxy is at a distance of 107 Mpc with an effective radius of 6.9 kpc. \cite{Montes:24} highlight that Nube is ten times fainter and has an effective radius three times larger than typical ultra diffuse galaxies with similar stellar masses. They are not able to reproduce the properties of Nube with current cold dark matter cosmological simulations which include baryonic feedback effects. {WALLABY J032705-211110 is also a very distant source, located at 157 Mpc. Its large size may also be a consequence of the galaxy-galaxy interactions taking place within the pair.}

 {As noted in Section \ref{sec:pair_phot}, if UDG-2 were to be observed face-on, its central surface brightness may be as faint as 26.3 mag arcsec$^{-2}$. Figure \ref{fig:hmu0} illustrates that while most detected UDGs have central surface brightnesses in between 24 mag arcsec$^{-2}$ and 25 mag arcsec$^{-2}$, there is still a tail extending to 26.4 mag arcsec$^{-2}$. It appears that a `face-on UDG-2' would place it on the threshold of the usual range of detected UDGs, bridging the gap between optically dark and LSB H{\sc i} sources. }

\section{Conclusions}
\label{sec:concl}

Using the WALLABY H{\sc i} data, we have identified a {core} sample of {10} UDGs defined by $\mu_{g,0}\ge24$ mag~arcsec$^{-2}$ and $R_{\rm eff}\ge1.5$ kpc \citep{vanDokkum:15}, and a broader sample including {12} additional faint diffuse galaxies (with $\mu_{g,0}\ge23.7$ mag~arcsec$^{-2}$ and $R_{\rm eff}\ge1.3$ kpc). We use the python package \textsc{AstroPhot} to create Sérsic models of the Legacy Survey images. Joint modelling of the $g$ and $r$-bands (or $i$-band where the $r$-band image is not available) allows for consistent and signal-to-noise optimised photometry. {All of the WALLABY UDGs exhibit high H{\sc i}-to-stellar mass ratios. The most extreme case is WALLABY J024013-393446, with a ratio of 27, while the sample as a whole has a mean ratio of 8.} 

Within our UDG sample, {not only do we have a UDG-LSBG pair,} we have also made the first discovery of a UDG-UDG pair, both of which possess H\,{\sc i}. The projected separation between the optical centres of the UDG-UDG pair is just 75 arcsec (equivalent to 22 kpc at a distance of 61.9 Mpc), with a central H\,{\sc i} velocity difference of 34 km s$^{-1}$.  Both galaxies meet the \cite{vanDokkum:15} UDG criteria: we measure the central $g$-band surface brightness and effective radius of the north-western UDG (UDG-1) to be 24.1 mag arcsec$^{-2}$ and 2.8 kpc, and the south-eastern UDG (UDG-2) to be 24.6 mag arcsec$^{-2}$ and 2.2 kpc.  UDG-1 has a larger H\,{\sc i} reservoir ($\log_{10}(M_{HI}/\rm M_{\odot})=8.95\pm0.03$) than UDG-2 ($\log_{10}(M_{HI}/\rm M_{\odot})=8.60\pm0.04$). Additionally, the stellar mass and SFR of UDG-1 ($\log_{10}(M_*/\rm M_{\odot}) = 8.07\pm0.12$, $\log_{10}(SFR/\rm M_{\odot}~yr^{-1}) = -1.27\pm0.12$) are approximately an order of magnitude larger than those of its companion ($\log_{10}(M_*/\rm M_{\odot}) = 7.16\pm0.12$,$\log_{10}(SFR/\rm M_{\odot}~yr^{-1}) = -2.10\pm0.12$).  Both UDGs demonstrate some evidence of a velocity gradient in their moment 1 and position-velocity diagrams, however, higher resolution data is required to confirm if they are rotation supported.

The UDG pair has an isolated local environment, with no other {known} galaxies or H\,{\sc i} sources within 30 arcmin (525 kpc at a distance of 61.9 Mpc) and $\pm1000$ km s$^{-1}$. The nearest galaxy is ESO 501-G082, with a projected separation of 30.3 arcmin and 240 km s$^{-1}$. However, in the context of the larger-scale structure, the pair is relatively less isolated. While it is located outside the virial radius of the Hydra cluster, the position on the phase-space diagram indicates {that} they are infalling into the cluster.  { We investigate the potential tidal interactions between the pair and the Hydra cluster as well as within the pair itself.  We explore the effect of using different methods to estimate the {rotational} velocities and dynamical masses (as we are unable to create accurate kinematic models with marginally resolved H{\sc i} data).  When adopting the larger dynamical masses derived from the bTF relation, the pair appears to be strongly gravitationally bound to each other, however, when using dynamical masses derived from the inclination corrected {$w_{20}$} emission line widths, the pair appears to be much less strongly bound. Despite this,} the pair does meet the merger selection criteria from \cite{Pfister:20}. Although thus far in its evolution, it is unlikely {that} the pair had been influenced by its environment (locally isolated), its future evolution may be impacted as the UDGs interact with each other and the Hydra cluster. 

{Currently, it is unclear whether isolated gas-rich UDGs are expected to lie on the bTF relation. Assuming the inclination corrected $w_{20}$ values are an accurate representation of the rotational velocity, the pair would need to be located at a distance approximately five times closer than that inferred from their H{\sc i} velocities in order to fall on the bTF. This discrepancy could indicate that the distance, inclination, or $w_{20}$ measurements are significantly incorrect, or that, like some other gas-rich UDGs \citep{mancerapina:19,Siljeg:24}, they genuinely lie off the bTF. Higher-resolution observations are therefore required to model the systems more robustly and place stronger constraints on their rotational velocities.}

The properties of both UDG-1 and UDG-2 differ from the UDGs already in the Hydra cluster. Like other HUDs, the WALLABY UDGs, including the pair, have bluer colours and larger effective radii. The axis ratio of UDG-2 is significantly smaller than all of axis ratios of the cluster UDGs, SMUDGes HUDs, and other WALLABY UDGs, suggesting that edge-on UDGs are unlikely to be detected without correcting for inclination. The UDG-pair and most of the WALLABY UDGs closely follow the ALFALFA stellar mass-H{\sc i} mass relation. This suggests that the UDG-pair and the majority of WALLABY UDGs are consistent with the broader H{\sc i}-selected galaxy population in the low stellar mass regime. The sSFR of the pair and most the WALLABY UDGs are also consistent with other HUDs. Some WALLABY UDGs have the SFE of typical H{\sc i}-selected galaxies, while others are inefficient at converting the H{\sc i} gas to stars. As UDG-1 and UDG-2 are consistent with the other HUDs, it appears that their presence in a pair has not significantly impacted their properties thus far in their evolution.

Optically-faint galaxies play a key role in hierarchical structure formation in the $\Lambda$CDM framework. UDGs are particularly intriguing due to their extreme sizes and low surface brightnesses, and because they have been shown to reside in range of dark matter halo sizes and be influenced by a complex interplay of internal and external processes. Even though UDGs span a range of properties and environments, this pair is truly a unique system. The identification of this pair of HUDs just outside the Hydra cluster underscores the need to investigate the role of environment in UDG formation and evolution. This system provides a link between the gas-rich UDGs in the field to their quiescent counterparts found in the cluster environment.


\section*{Acknowledgements}


This scientific work uses data obtained from Inyarrimanha Ilgari Bundara / the Murchison Radio-astronomy Observatory. We acknowledge the Wajarri Yamaji People as the Traditional Owners and native title holders of the Observatory site. CSIRO’s ASKAP radio telescope is part of the Australia Telescope National Facility (https://ror.org/05qajvd42). Operation of ASKAP is funded by the Australian Government with support from the National Collaborative Research Infrastructure Strategy. ASKAP uses the resources of the Pawsey Supercomputing Research Centre. Establishment of ASKAP, Inyarrimanha Ilgari Bundara, the CSIRO Murchison Radio-astronomy Observatory and the Pawsey Supercomputing Research Centre are initiatives of the Australian Government, with support from the Government of Western Australia and the Science and Industry Endowment Fund.

WALLABY acknowledges technical support from the Australian SKA Regional Centre (AusSRC).

Parts of this research were supported by the Australian Research Council Centre of Excellence for All Sky Astrophysics in 3 Dimensions (ASTRO 3D), through project number CE170100013. 

This investigation has made use of the NASA/IPAC Extragalactic Database (NED) which is operated by the Jet Propulsion Laboratory, California Institute of Technology, under contract with the National Aeronautics and Space Administration, and NASA's Astrophysics Data System.

The Legacy Surveys consist of three individual and complementary projects: the Dark Energy Camera Legacy Survey (DECaLS; Proposal ID \#2014B-0404; PIs: David Schlegel and Arjun Dey), the Beijing-Arizona Sky Survey (BASS; NOAO Prop. ID \#2015A-0801; PIs: Zhou Xu and Xiaohui Fan), and the Mayall z-band Legacy Survey (MzLS; Prop. ID \#2016A-0453; PI: Arjun Dey). DECaLS, BASS and MzLS together include data obtained, respectively, at the Blanco telescope, Cerro Tololo Inter-American Observatory, NSF’s NOIRLab; the Bok telescope, Steward Observatory, University of Arizona; and the Mayall telescope, Kitt Peak National Observatory, NOIRLab. Pipeline processing and analyses of the data were supported by NOIRLab and the Lawrence Berkeley National Laboratory (LBNL). The Legacy Surveys project is honored to be permitted to conduct astronomical research on Iolkam Du’ag (Kitt Peak), a mountain with particular significance to the Tohono O’odham Nation.

NOIRLab is operated by the Association of Universities for Research in Astronomy (AURA) under a cooperative agreement with the National Science Foundation. LBNL is managed by the Regents of the University of California under contract to the U.S. Department of Energy.

This project used data obtained with the Dark Energy Camera (DECam), which was constructed by the Dark Energy Survey (DES) collaboration. Funding for the DES Projects has been provided by the U.S. Department of Energy, the U.S. National Science Foundation, the Ministry of Science and Education of Spain, the Science and Technology Facilities Council of the United Kingdom, the Higher Education Funding Council for England, the National Center for Supercomputing Applications at the University of Illinois at Urbana-Champaign, the Kavli Institute of Cosmological Physics at the University of Chicago, Center for Cosmology and Astro-Particle Physics at the Ohio State University, the Mitchell Institute for Fundamental Physics and Astronomy at Texas A\&M University, Financiadora de Estudos e Projetos, Fundacao Carlos Chagas Filho de Amparo, Financiadora de Estudos e Projetos, Fundacao Carlos Chagas Filho de Amparo a Pesquisa do Estado do Rio de Janeiro, Conselho Nacional de Desenvolvimento Cientifico e Tecnologico and the Ministerio da Ciencia, Tecnologia e Inovacao, the Deutsche Forschungsgemeinschaft and the Collaborating Institutions in the Dark Energy Survey. The Collaborating Institutions are Argonne National Laboratory, the University of California at Santa Cruz, the University of Cambridge, Centro de Investigaciones Energeticas, Medioambientales y Tecnologicas-Madrid, the University of Chicago, University College London, the DES-Brazil Consortium, the University of Edinburgh, the Eidgenossische Technische Hochschule (ETH) Zurich, Fermi National Accelerator Laboratory, the University of Illinois at Urbana-Champaign, the Institut de Ciencies de l’Espai (IEEC/CSIC), the Institut de Fisica d'Altes Energies, Lawrence Berkeley National Laboratory, the Ludwig Maximilians Universitat Munchen and the associated Excellence Cluster Universe, the University of Michigan, NSF’s NOIRLab, the University of Nottingham, the Ohio State University, the University of Pennsylvania, the University of Portsmouth, SLAC National Accelerator Laboratory, Stanford University, the University of Sussex, and Texas A\&M University.

BASS is a key project of the Telescope Access Program (TAP), which has been funded by the National Astronomical Observatories of China, the Chinese Academy of Sciences (the Strategic Priority Research Program “The Emergence of Cosmological Structures” Grant \# XDB09000000), and the Special Fund for Astronomy from the Ministry of Finance. The BASS is also supported by the External Cooperation Program of Chinese Academy of Sciences (Grant \# 114A11KYSB20160057), and Chinese National Natural Science Foundation (Grant \# 12120101003, \# 11433005).

The Legacy Survey team makes use of data products from the Near-Earth Object Wide-field Infrared Survey Explorer (NEOWISE), which is a project of the Jet Propulsion Laboratory/California Institute of Technology. NEOWISE is funded by the National Aeronautics and Space Administration.

The Legacy Surveys imaging of the DESI footprint is supported by the Director, Office of Science, Office of High Energy Physics of the U.S. Department of Energy under Contract No. DE-AC02-05CH1123, by the National Energy Research Scientific Computing Center, a DOE Office of Science User Facility under the same contract; and by the U.S. National Science Foundation, Division of Astronomical Sciences under Contract No. AST-0950945 to NOAO.

NA would like to acknowledge funding from the Canada First Research Excellence Fund through the Arthur B. McDonald Canadian Astroparticle Physics Research Institute.
KS acknowledges support from the Natural Sciences and Engineering Research Council of Canada (NSERC).
PEMP acknowledges the support from the Dutch Research Council (NWO) through the Veni grant VI.Veni.222.364.

\section*{Data Availability}

The WALLABY source catalogue and associated data products (e.g. cubelets, moment maps, integrated spectra, radial surface density profiles) are available online through the CSIRO ASKAP Science Data Archive (CASDA) and the Canadian Astronomy Data Centre (CADC). All source and kinematic model data products are mirrored at both locations. Links to the data access services and the software tools used to produce the data products as well as documented instructions and example scripts for accessing the data are available from the WALLABY Data Portal (\url{https://wallaby-survey.org/data/}).



\bibliographystyle{mnras}
\bibliography{example} 




\appendix

\section{SMUDGes sources identified in WALLABY}
\label{sec:appen_smudges}

{ Table \ref{tab:smudges} presents the SMUDGes names associated with the corresponding WALLABY sources as discussed in Section \ref{sec:smudges}. }

\begin{table}
    \centering
    \begin{tabular}{cc}
    \hline
       WALLABY  &   SMUDGes \\
       \hline
        WALLABY J205043-531619 & SMDG2050426-531625 \\
        WALLABY J032705-211110 & SMDG0327071-211103 \\
        WALLABY J033402-205037 & SMDG0334020-205042 \\
        WALLABY J125311+032639 & SMDG1253110+032631 \\
        WALLABY J125603+045202 & SMDG1256034+045202 \\
        WALLABY J010649-601631 & SMDG0106499-601634 \\
        WALLABY J033408-232125 & SMDG0334081-232128 \\
        WALLABY J125604+034843 & SMDG1256044+034846 \\
        WALLABY J124232-012111 & SMDG1242255-012026 \\
        WALLABY J125313+042746 & SMDG1253107+042638 \\
        \hline
    \end{tabular}
    \caption{The SMUDGes names associated with the corresponding WALLABY sources.}
    \label{tab:smudges}
\end{table}

\section{The UDG sample figures}
\label{sec:appen_figs}

This section contains the 3 colour optical images with H{\sc i} contours overlaid as well as the moment 1 maps for the {combined core and broad} UDG sample (excluding the pair). 


\begin{figure}
    \centering
    \includegraphics[width=\linewidth]{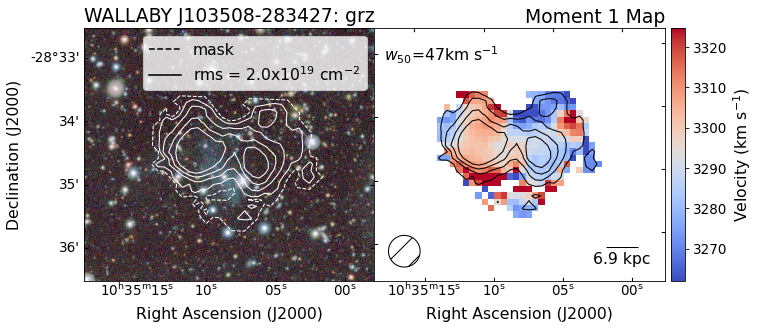}
    \caption{ }
\end{figure}

\begin{figure}
    \centering
    \includegraphics[width=\linewidth]{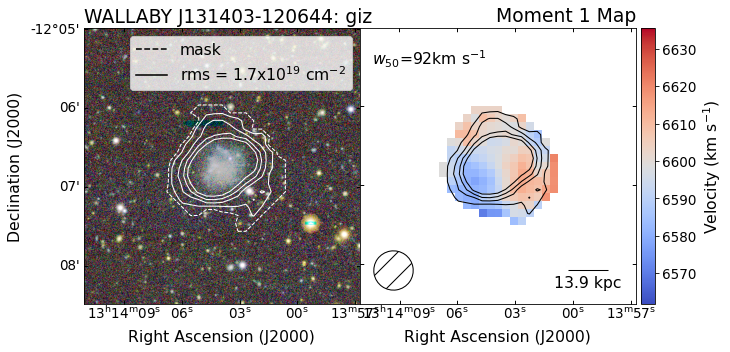}
    \caption{ }
    \label{fig:a3}
\end{figure}

\begin{figure}
    \centering
    \includegraphics[width=\linewidth]{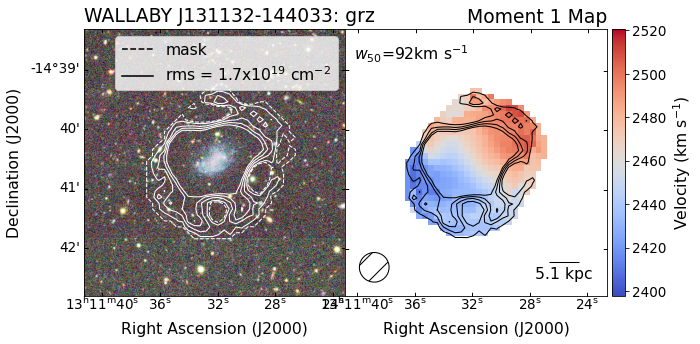}
    \caption{ }
\end{figure}

\begin{figure}
    \centering
    \includegraphics[width=\linewidth]{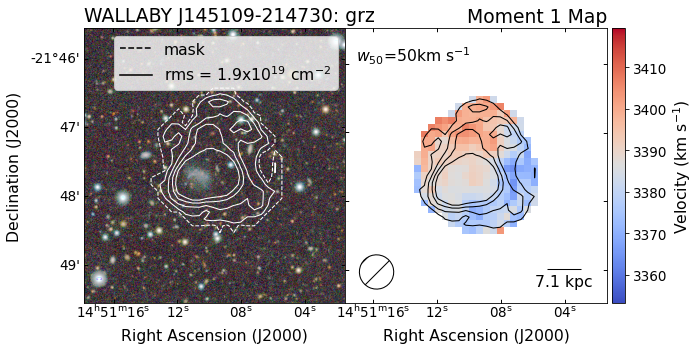}
    \caption{ }
\end{figure}

\begin{figure}
    \centering
    \includegraphics[width=\linewidth]{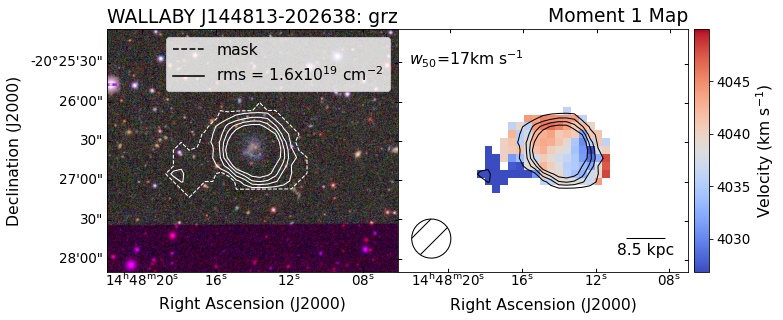}
    \caption{ }
\end{figure}

\begin{figure}
    \centering
    \includegraphics[width=\linewidth]{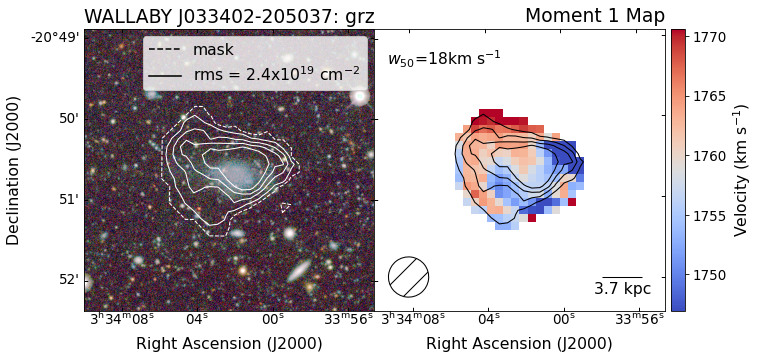}
    \caption{ }
\end{figure}

\begin{figure}
    \centering
    \includegraphics[width=\linewidth]{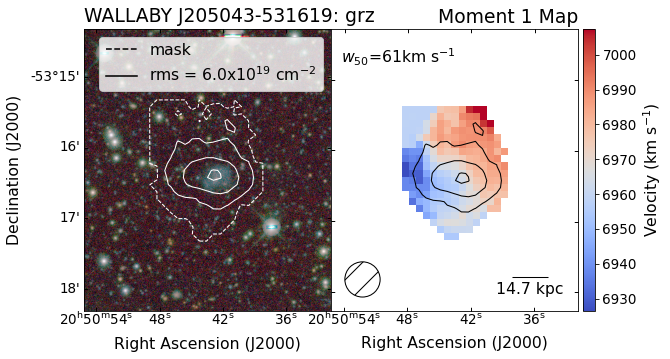}
    \caption{ }
\end{figure}

\begin{figure}
    \centering
    \includegraphics[width=\linewidth]{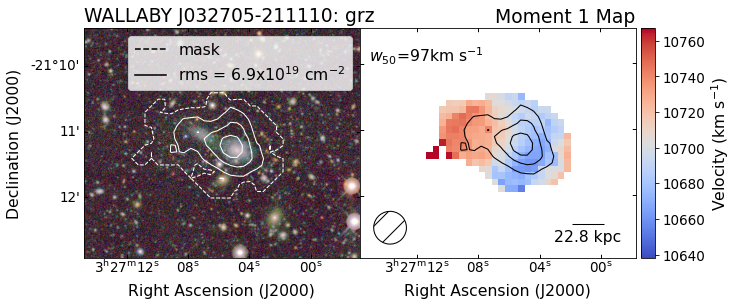}
    \caption{ }
\end{figure}

\begin{figure}
    \centering
    \includegraphics[width=\linewidth]{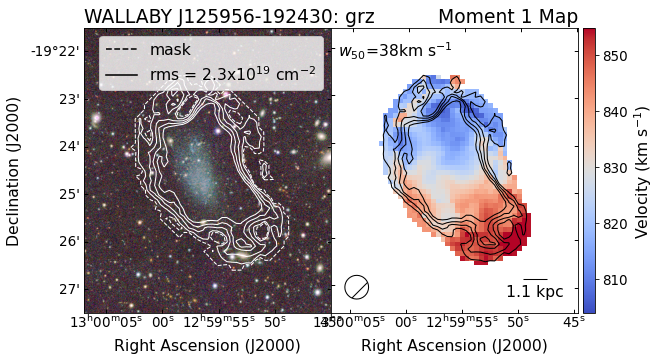}
    \caption{ }
\end{figure}

\begin{figure}
    \centering
    \includegraphics[width=\linewidth]{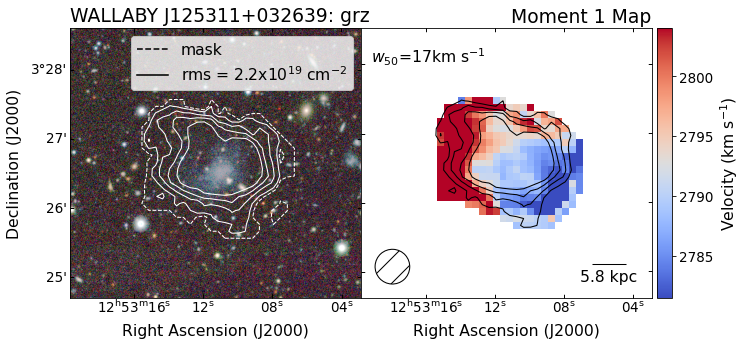}
    \caption{ }
\end{figure}

\begin{figure}
    \centering
    \includegraphics[width=\linewidth]{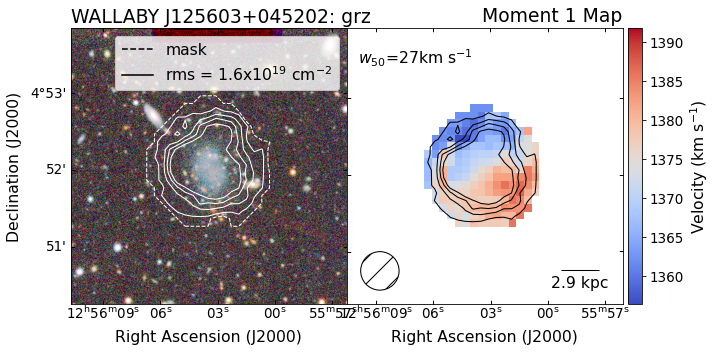}
    \caption{ }
\end{figure}

\begin{figure}
    \centering
    \includegraphics[width=\linewidth]{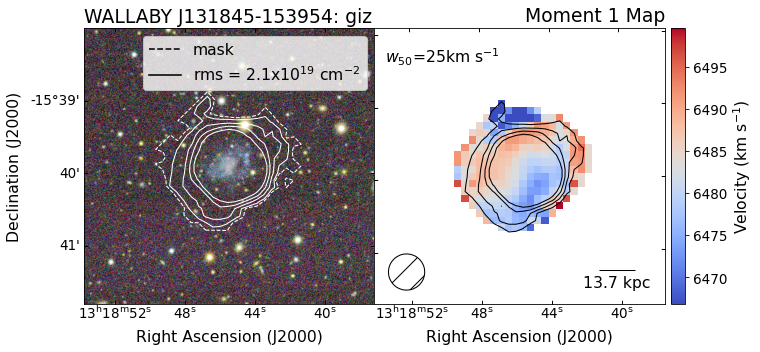}
    \caption{ }
\end{figure}

\begin{figure}
    \centering
    \includegraphics[width=\linewidth]{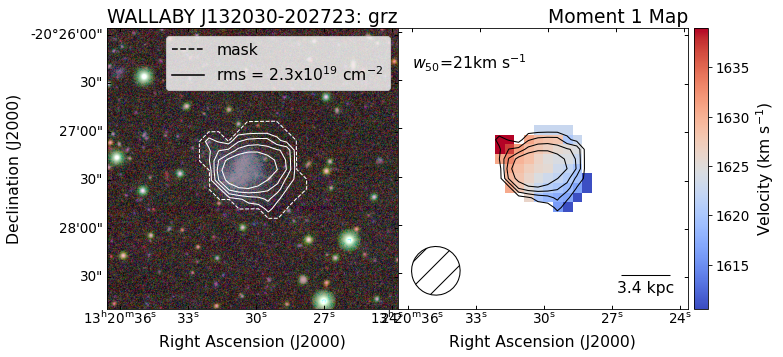}
    \caption{ }
\end{figure}

\begin{figure}
    \centering
    \includegraphics[width=\linewidth]{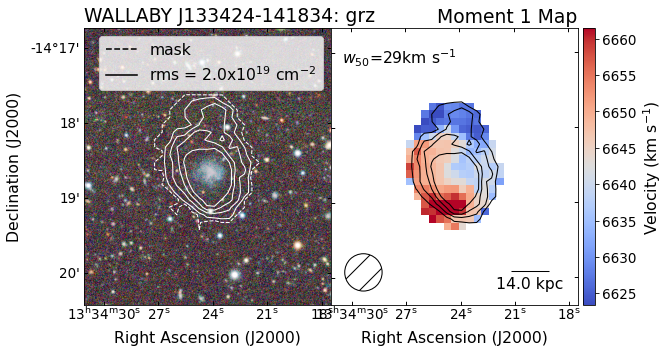}
    \caption{ }
\end{figure}

\begin{figure}
    \centering
    \includegraphics[width=\linewidth]{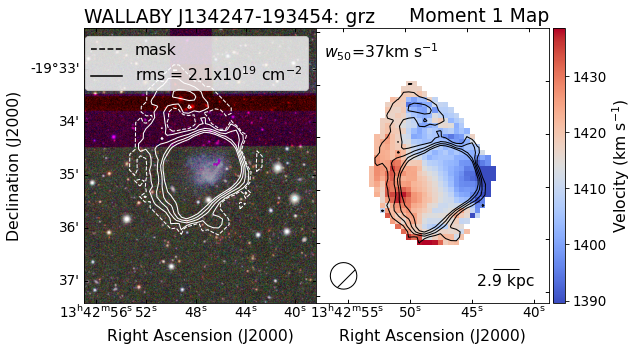}
    \caption{ }
\end{figure}

\begin{figure}
    \centering
    \includegraphics[width=\linewidth]{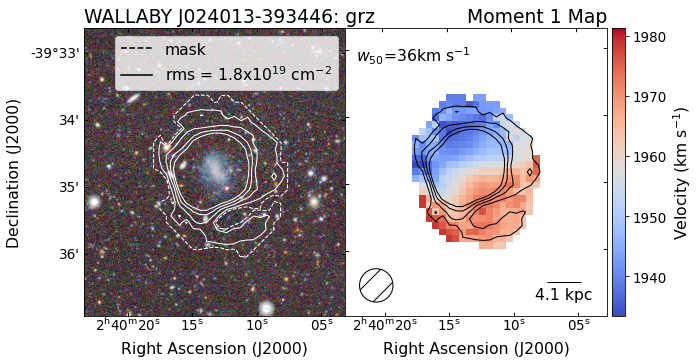}
    \caption{ }
\end{figure}

\begin{figure}
    \centering
    \includegraphics[width=\linewidth]{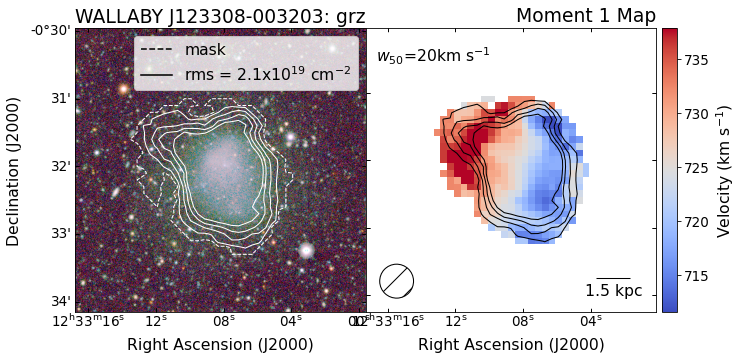}
    \caption{ }
\end{figure}

\begin{figure}
    \centering
    \includegraphics[width=\linewidth]{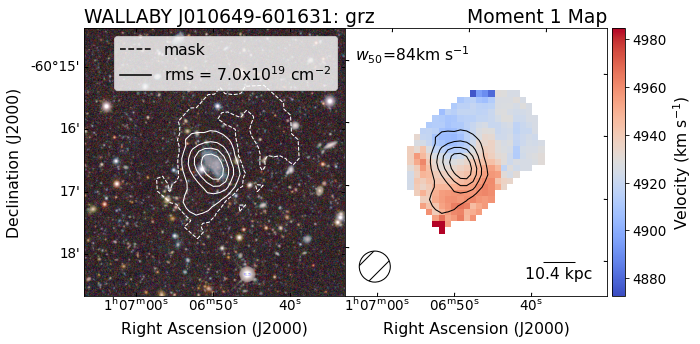}
    \caption{ }
\end{figure}

\begin{figure}
    \centering
    \includegraphics[width=\linewidth]{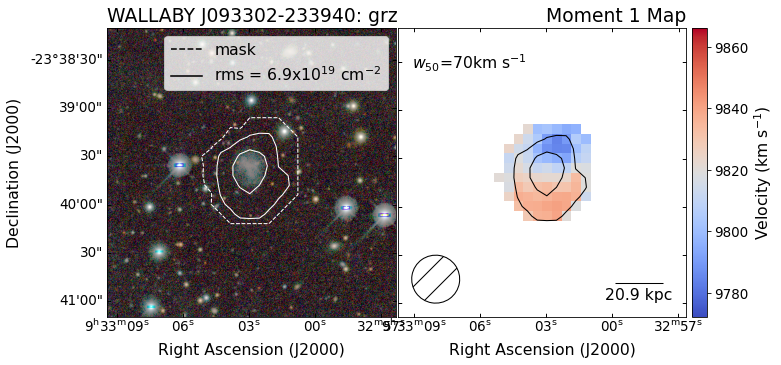}
    \caption{ }
\end{figure}

\begin{figure}
    \centering
    \includegraphics[width=\linewidth]{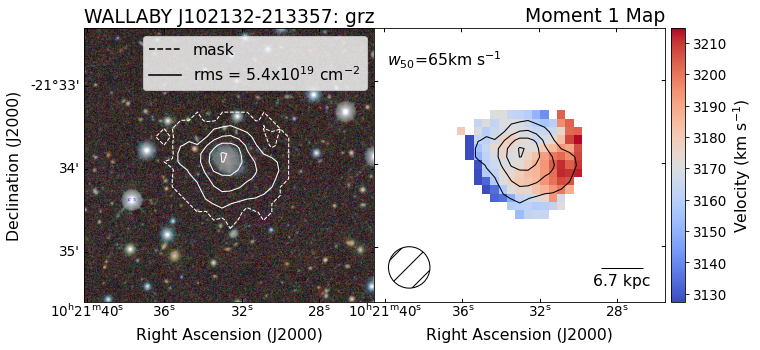}
    \caption{ }
\end{figure}

\section{Photometric properties}
\label{sec:append_models}

\begin{figure}
     \centering
       \begin{subfigure}[t]{0.35\textwidth}
         \centering
         \includegraphics[width=\textwidth]{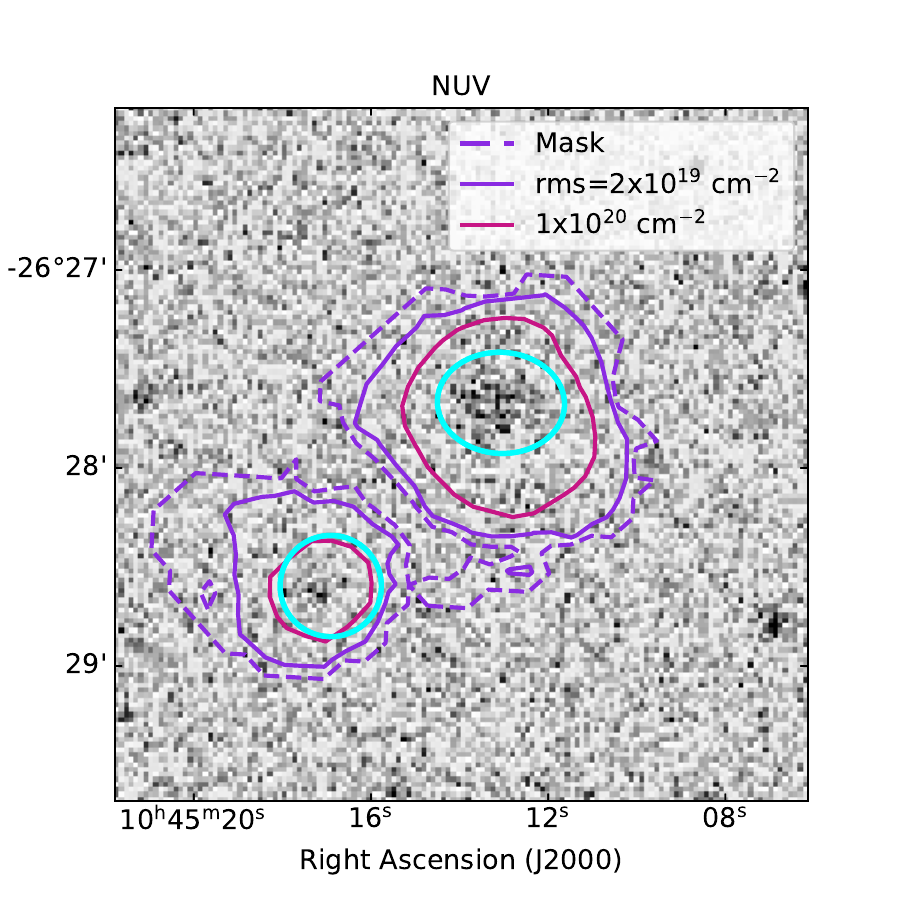}
         \caption{  }
         \label{fig:nuv}
     \end{subfigure}
     \begin{subfigure}[t]{0.35\textwidth}
         \centering
         \includegraphics[width=\textwidth]{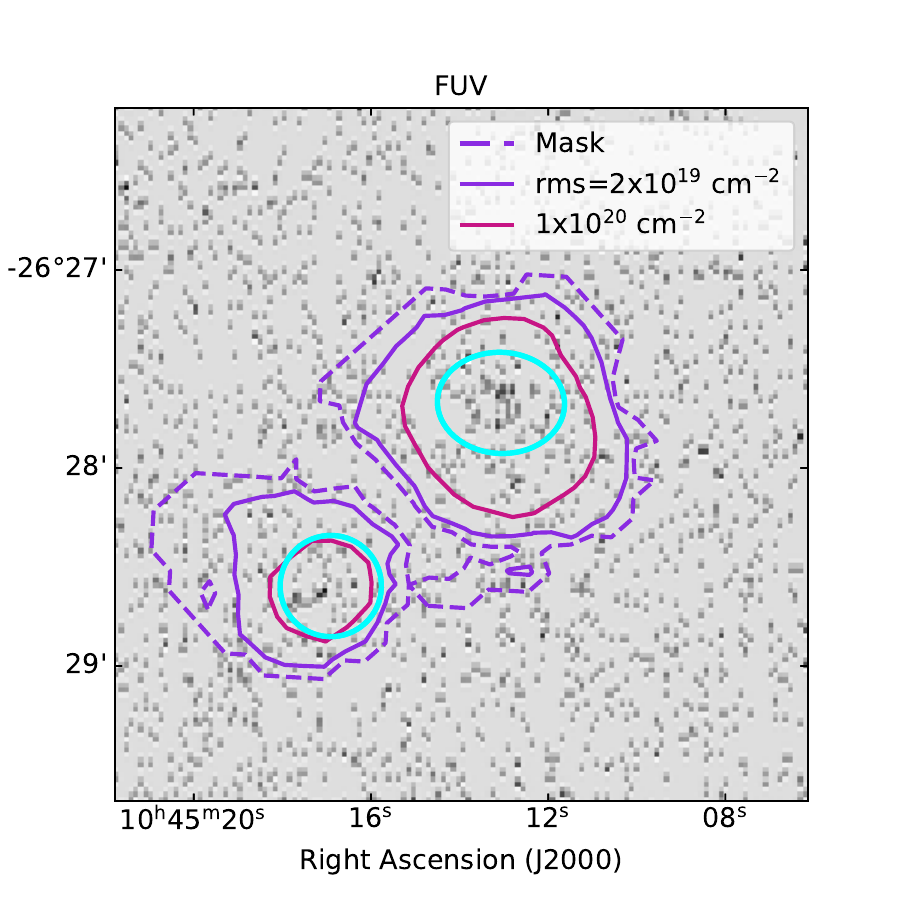}
         \caption{  }
         \label{fig:fuv}
     \end{subfigure}
    \caption{ GALEX images with H\,{\sc i} contours ({purple and pink) and aperture (cyan) overlaid. (a) NUV GALEX image, (b) FUV GALEX image. The dashed line corresponds to the edge of the SoFiA mask, and the lowest contour level shown represents the local rms. The apertures have the central coordinates and position angle from the Sérsic models and a radius equal to $2R_e$. UDG-1's aperture has the axis ratio from the Sérsic model, however a circlar aperture is used for UDG-2 as discussed in Section \ref{sec:photometry}.}}
    \label{fig:galex}
\end{figure}

\begin{figure}
     \centering
       \begin{subfigure}[t]{0.28\textwidth}
         \centering
         \includegraphics[width=\textwidth]{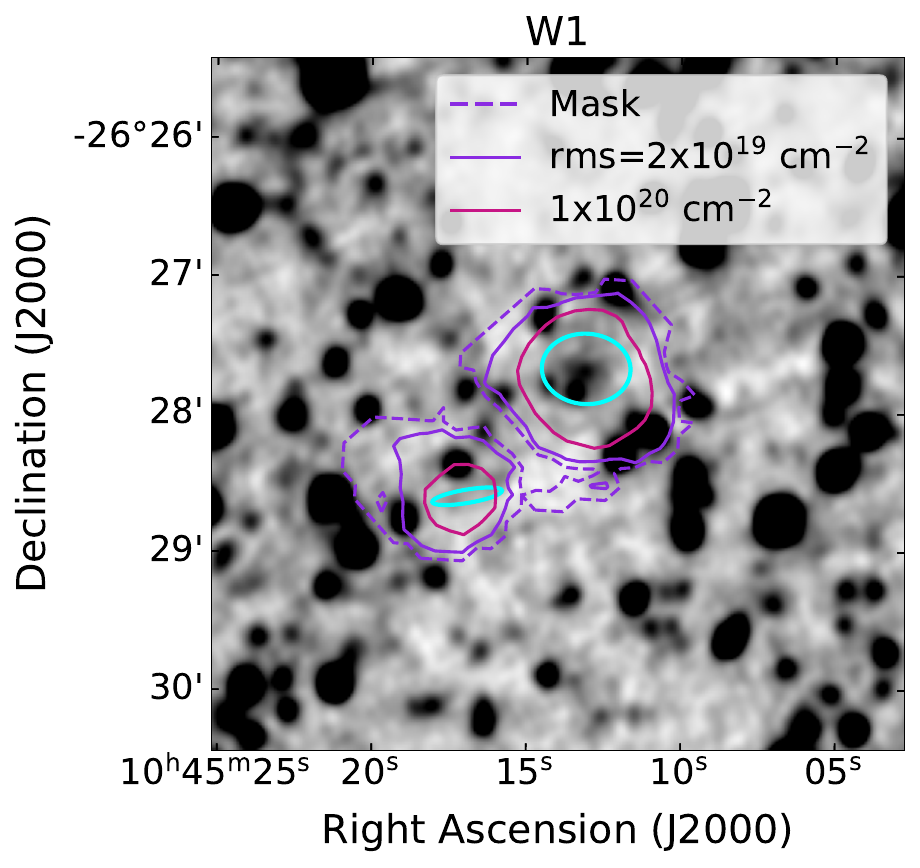}
         \caption{  }
         \label{fig:w1}
     \end{subfigure}
     \begin{subfigure}[t]{0.28\textwidth}
         \centering
         \includegraphics[width=\textwidth]{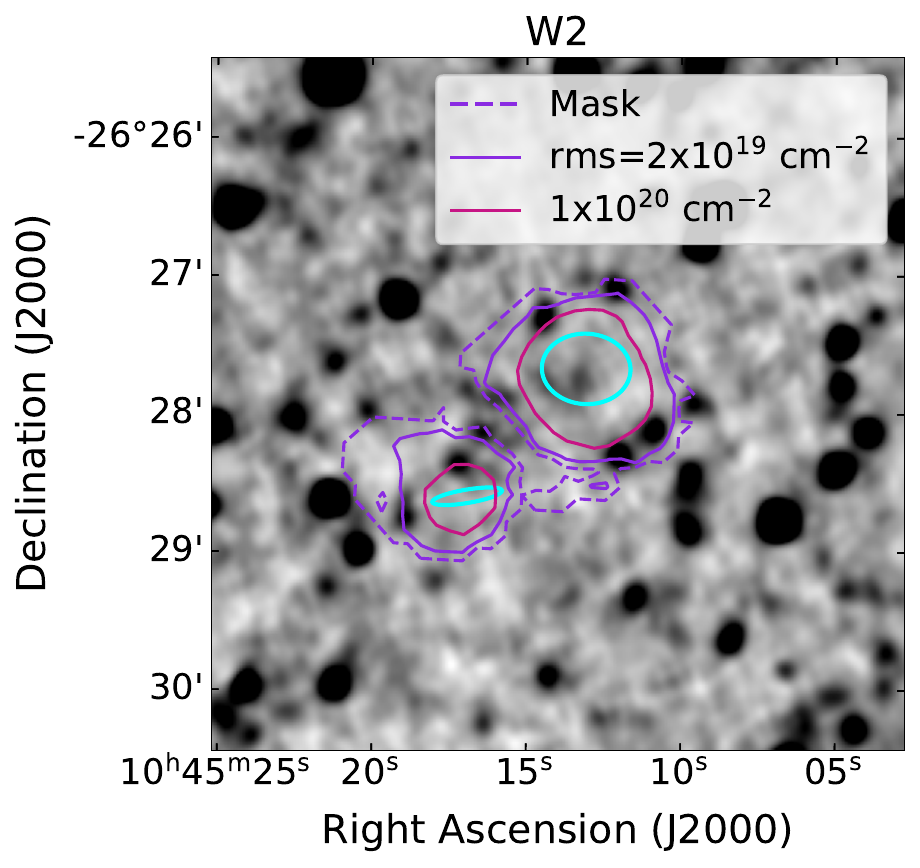}
         \caption{  }
         \label{fig:w2}
     \end{subfigure}
     \begin{subfigure}[b]{0.28\textwidth}
         \centering
         \includegraphics[width=\textwidth]{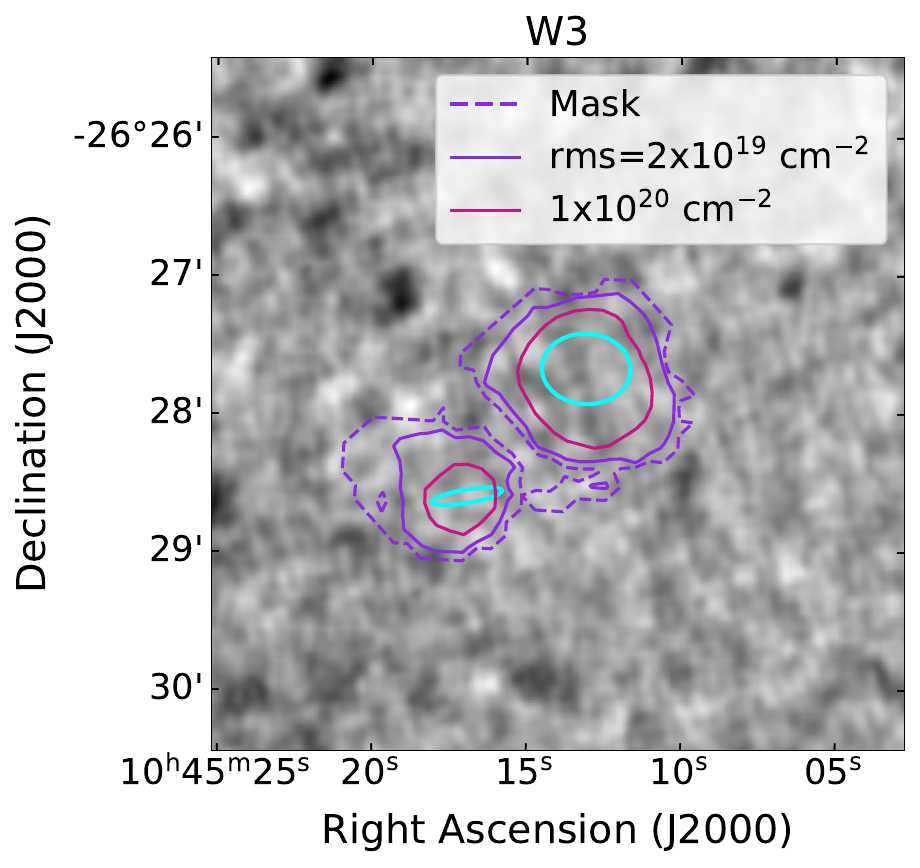}
         \caption{  }
         \label{fig:w3}
     \end{subfigure}
    \begin{subfigure}[b]{0.28\textwidth}
         \centering
         \includegraphics[width=\textwidth]{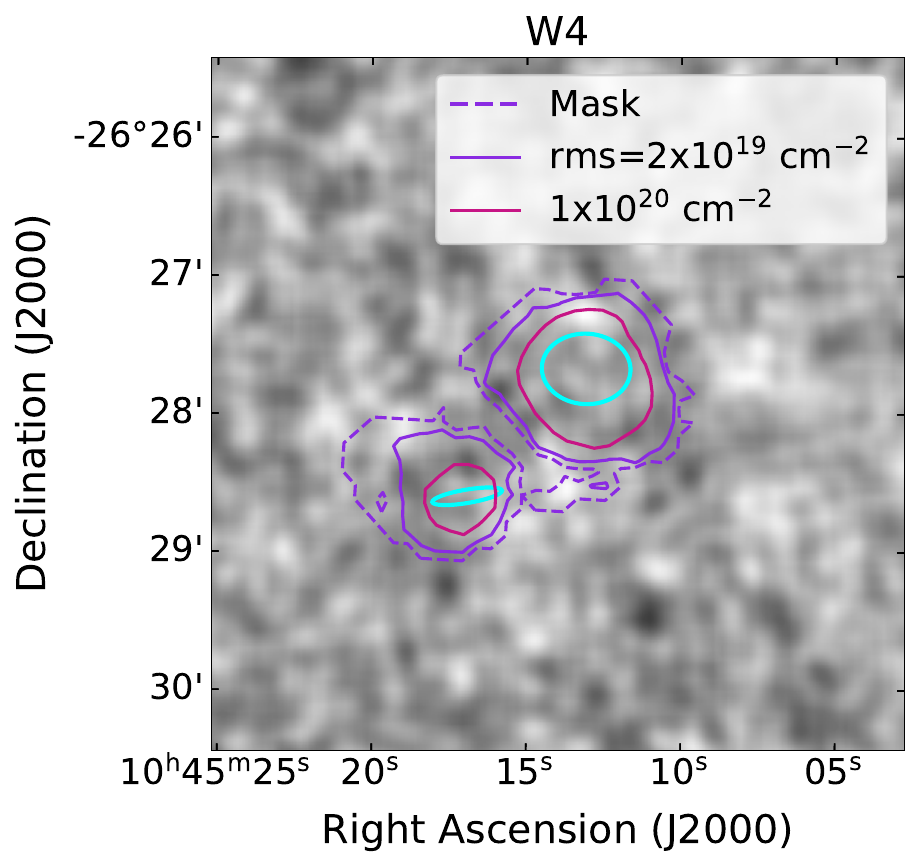}
         \caption{  }
         \label{fig:w4}
     \end{subfigure}
     
    \caption{ WISE images of WALLABY J104513-262755 with H\,{\sc i} contours and optical apertures from the Sérsic fits with radius $=2R_e$ overlaid. (a) W1 WISE image, (b) W2 WISE image, (c) W3 WISE image, (d) W4 WISE image. UDG-1 has a faint counterpart visible in the W1 and W2 images, while UDG-2 has no visible counterpart in any of the WISE bands.}
    \label{fig:wise}
\end{figure}

In this section we present additional figures referred to in Section \ref{sec:pair_phot}. {Figure \ref{fig:galex} shows the NUV and FUV GALEX images with H\,{\sc i} contours and optical apertures overlaid (with radius = $2R_e$, note a circular aperture is used here for UDG-2 as discussed in Section \ref{sec:photometry}).} Figure \ref{fig:wise} shows the WISE images with H\,{\sc i} contours and optical apertures with radius $=2R_e$ overlaid. UDG-1 is only detected in the W1 and W2 bands, and UDG-2 is not detected at all any of the four WISE bands.


\newpage
\section{Interactions}
\label{appen:int}

Here we include the Figures that help to visualise for which projection angles the UDG pair are gravitationally bound to each other and the Hydra cluster following \protect\cite{Wolfinger:16}, as discussed in Section \ref{sec:int}.

\begin{figure}
    \centering
     
     \begin{subfigure}[b]{0.35\textwidth}
         \centering
         \includegraphics[width=\textwidth]{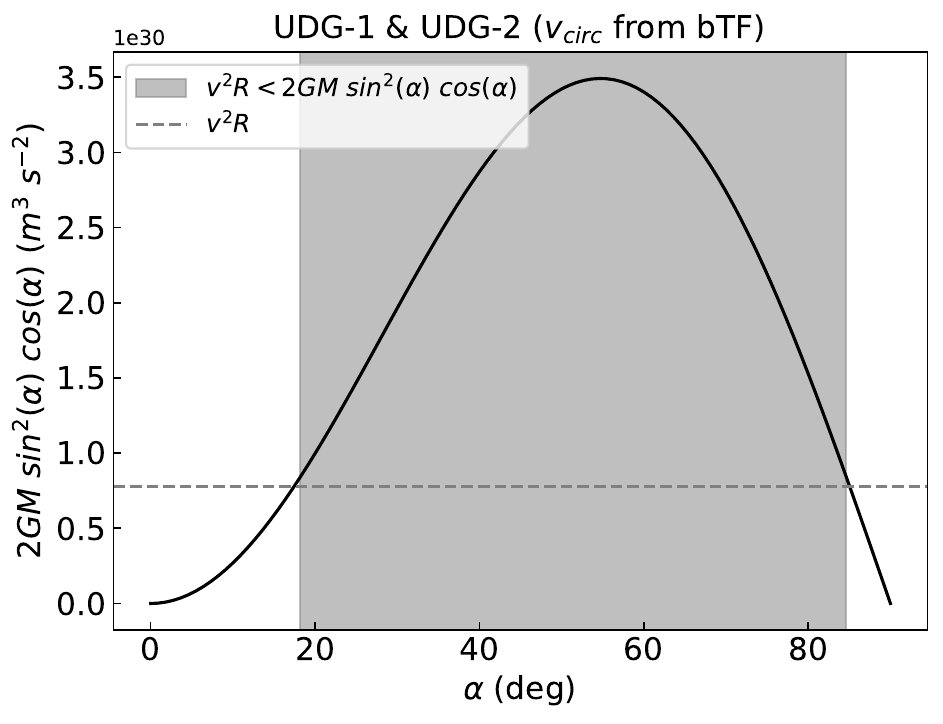}
         \caption{ }
         \label{fig:12TF}
     \end{subfigure}
     \begin{subfigure}[b]{0.34\textwidth}
         \centering
         \includegraphics[width=\textwidth]{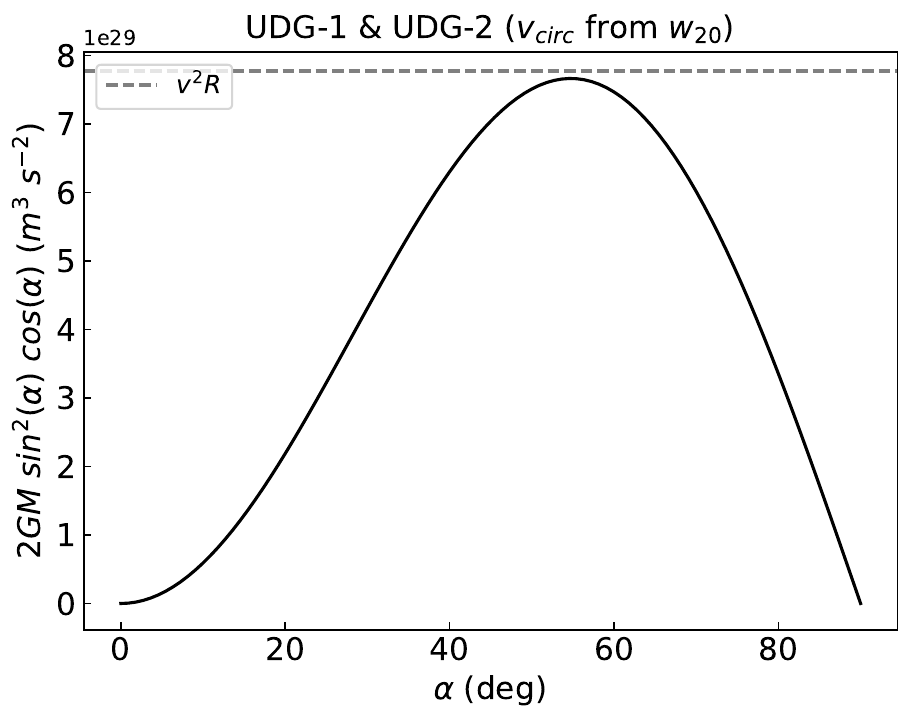}
         \caption{ }
         \label{fig:12w50}
     \end{subfigure}
    \vfill
     \begin{subfigure}[t]{0.35\textwidth}
         \centering
         \includegraphics[width=\textwidth]{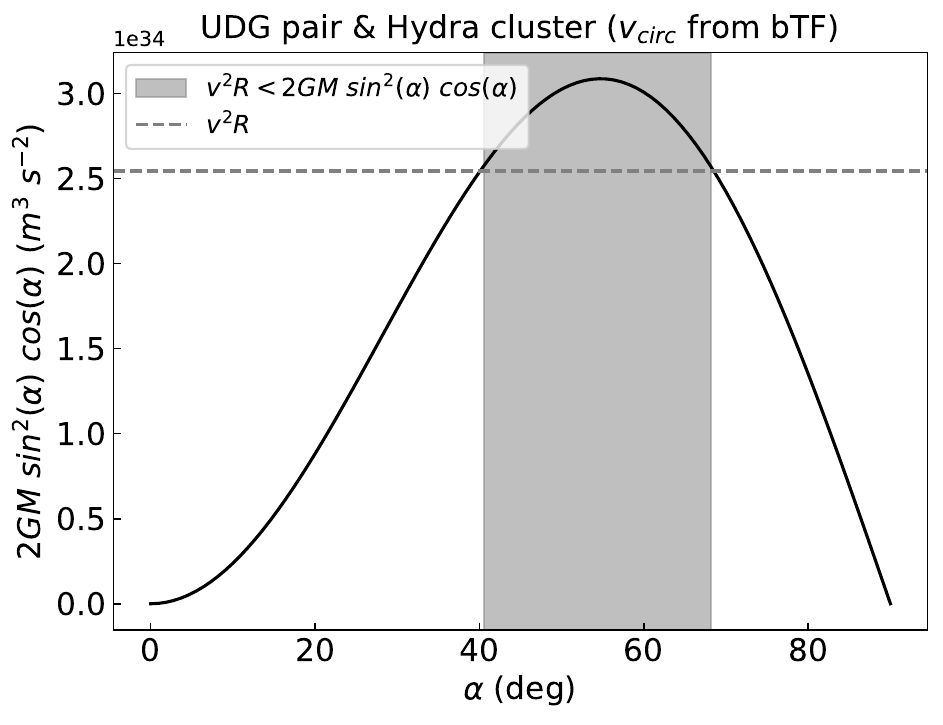}
         \caption{ }
         \label{fig:pcTF}
     \end{subfigure}
     \begin{subfigure}[t]{0.35\textwidth}
         \centering
         \includegraphics[width=\textwidth]{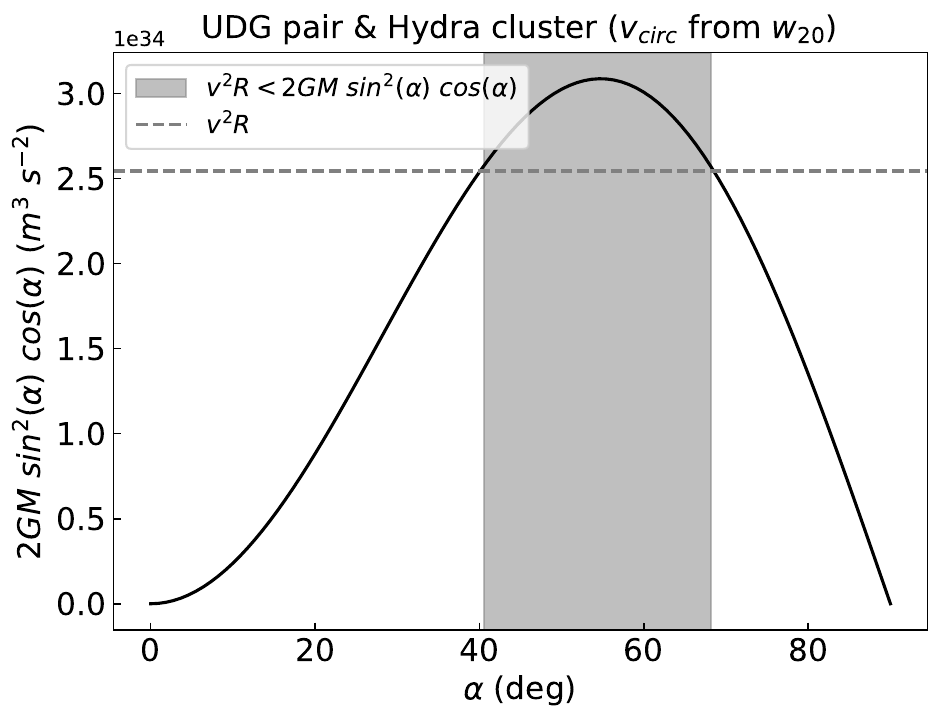}
         \caption{ }
         \label{fig:pcw50}
     \end{subfigure}
     
 \caption{A two body system is gravitationally bound if the kinetic energy is less or equal to the potential energy of the system. Following \protect\cite{Wolfinger:16}, the system is bound if Equation \ref{eq:bound} is satisfied. The shaded regions show for which projection angles $\alpha$ the UDG pair are bound to each other/the Hydra cluster. (a) UDG-1 and UGD-2 ($v_{circ}$ from the bTF relation. (b) UDG-1 and UGD-2 ($v_{circ}$ from $w_{20}$). (c) The UDG pair and the Hydra cluster ($v_{circ}$ from the bTF relation). (d) The UDG pair and the Hydra cluster ($v_{circ}$ from $w_{20}$).}
    \label{fig:int}
\end{figure}


\bsp	
\label{lastpage}
\end{document}